\documentclass[a4paper,11pt]{article}
\usepackage{jheppub} 
\usepackage{graphicx}
\usepackage{subcaption}
\usepackage{amsmath}
\usepackage{subcaption}
\usepackage{bm}
\makeatletter
\def\@fpheader{}
\def\@preprint{}
\makeatother
\title{\boldmath Thermodynamics and transport in holographic QCD with Gauss–Bonnet corrections}

\author[a]{ChenWei Tong,}
\author[a]{Jie Zhou,}
\author[a]{YuanXu Wang,}
\author[b,1]{Rong-Gen Cai\note{Corresponding author.},}
\author[b, a, 1]{Song He,}
\author[c,d,e,1]{Li Li}

\affiliation[a]{Center for Theoretical Physics and College of Physics, Jilin University, Changchun 130012, China}

\affiliation[b]{Institute of Fundamental Physics and Quantum Technology, Ningbo University, Ningbo, Zhejiang 315211, China}

\affiliation[c]{Institute of Theoretical Physics, Chinese Academy of Sciences, Beijing 100190, China}

\affiliation[d]{School of Physical Sciences, University of Chinese Academy of Sciences, Beijing 100049, China}
\affiliation[e]{School of Fundamental Physics and Mathematical Sciences, Hangzhou Institute for Advanced Study, UCAS, Hangzhou
310024, China}

\emailAdd{tongcw24@mails.jlu.edu.cn}
\emailAdd{zhoujie23@ciac.ac.cn}
\emailAdd{yuanxu23@mails.jlu.edu.cn}
\emailAdd{caironggen@nbu.edu.cn}
\emailAdd{hesong@nbu.edu.cn}
\emailAdd{liliphy@itp.ac.cn}

\abstract{Thermodynamics and transport are investigated in a holographic QCD model that extends the Einstein--Maxwell--Dilaton framework by incorporating Gauss--Bonnet corrections. Model parameters are fixed using state-of-the-art lattice QCD thermodynamics. The analysis then examines the equation of state at zero and finite baryon chemical potential, the phase structure in the temperature and chemical potential plane, as well as the shear and bulk viscosity to entropy ratios, \(\eta/s\) and \(\zeta/s\),  via the corresponding fluctuation equations. For a constant Gauss--Bonnet coupling, the model preserves a reasonable description of the equation of state and generates a temperature-dependent \(\eta/s\), although the resulting profile remains monotonic near the crossover region, which does not satisfy the phenomenological expectation. When the Gauss--Bonnet coupling is allowed to depend on the dilaton, a non-monotonic \(\eta/s\) and a peaked \(\zeta/s\) are obtained while maintaining agreement with thermodynamic constraints. The resulting phase diagram contains a critical end point in a phenomenologically relevant region.}

\begin{document}
	\maketitle

\section{Introduction}
\label{sec:intro}
Quantum Chromodynamics (QCD), the theory of the strong interaction, is asymptotically free in the ultraviolet and strongly coupled in the infrared \cite{Gross:1973id, Politzer:1973fx, Wilson:1974sk}. As a result, while perturbative methods provide an accurate description of hard processes, the low-energy regime of QCD is governed by genuinely non-perturbative phenomena, including confinement, chiral symmetry breaking, and the emergence of strongly interacting matter under extreme conditions. A central open problem in this context is the structure of the QCD phase diagram at finite temperature $T$ and baryon chemical potential $\mu_B$, in particular, the possible existence of a critical end point (CEP) \cite{Bzdak:2019pkr,Pandav:2022xxx,Pasztor:2024dpv}. If present, the CEP separates the first-order transition line at large baryon density from the crossover region at lower density. Its existence and location are therefore of direct relevance for the thermodynamics of QCD matter. However, a first-principles determination of the CEP remains difficult, largely because lattice QCD at finite baryon chemical potential is obstructed by the sign problem \cite{Nagata:2021ugx}. This makes heavy-ion collisions a natural setting in which to search for possible critical behavior, with fluctuations of conserved charges providing particularly sensitive observables \cite{Asakawa:2015ybt,An:2021wof,Adel:2017uxi}.

Heavy-ion collision experiments at RHIC and the LHC also provide access to the strongly coupled quark--gluon plasma (QGP) produced at high temperature \cite{Luzum:2008cw}. A key result from these experiments is that the QGP behaves as a nearly perfect fluid with strong collective flow. In particular, the elliptic flow coefficient \(v_2\) in non-central collisions is highly sensitive to the shear viscosity of the medium \cite{Luzum:2010ag,Nagle:2011uz,Teaney:2003kp,Borsanyi:2010bp}. This observation has established relativistic hydrodynamics as the standard effective description of the spacetime evolution of the plasma, with transport coefficients such as the shear and bulk viscosities playing a central role. Their computation from first principles, however, remains challenging. Although lattice QCD successfully determines many equilibrium properties of QCD matter near vanishing baryon chemical potential \cite{Borsanyi:2013bia,HotQCD:2014kol,Philipsen:2012nu}, transport coefficients are not directly accessible in Euclidean simulations. Their extraction requires analytic continuation of correlation functions, which is an ill-posed inverse problem and leads to substantial uncertainties \cite{Meyer:2007dy, Song:2010mg}. At finite baryon density, the situation is further complicated by the sign problem, which severely limits standard lattice Monte Carlo methods.

The strong collective flow observed in the QGP indicates that the shear viscosity to entropy density ratio \(\eta/s\) is small and may approach the Kovtun--Son--Starinets bound, \(\eta/s=1/(4\pi)\), predicted from holographic theories at strong coupling \cite{Policastro:2001yc, Kovtun:2003wp}. The temperature dependence of \(\eta/s\) is of particular interest, since this ratio is expected to develop a minimum near the transition temperature \(T_c\) and thus provides a sensitive probe of the QCD crossover \cite{Cremonini:2012ny}. Away from the transition, \(\eta/s\) is expected to increase both in the hadronic phase, where effective hadronic interactions become less efficient \cite{Gavin:1985ph,Prakash:1993bt}, and in the high-temperature deconfined regime, where asymptotic freedom becomes operative. A first-principles determination of this behavior remains difficult because transport coefficients are real-time observables: although they can be expressed through Kubo formulas in terms of retarded stress-tensor correlators, their extraction from Euclidean lattice data requires analytic continuation and is therefore subject to large uncertainties. Recent Bayesian analyses have nevertheless provided increasingly quantitative constraints on \(\eta/s(T)\) \cite{JETSCAPE:2020shq}, with the JETSCAPE framework~\cite{JETSCAPE:2020mzn} in particular pointing to a nontrivial temperature dependence. The bulk viscosity \(\zeta\) is even more sensitive to the nonconformal dynamics of QCD and is expected to be enhanced near the deconfinement transition.

Holographic methods based on gauge/gravity duality provide a useful framework for studying strongly coupled QCD matter by relating higher-dimensional gravitational dynamics to observables in the boundary gauge theory \cite{Maldacena:1997re,Witten:1998qj,Gubser:1998bc,Witten:1998zw}. In particular, Einstein--Maxwell--Dilaton (EMD) models with suitably chosen scalar potentials and gauge couplings have been shown to reproduce lattice-QCD thermodynamics with good accuracy, and have been widely used to investigate the QCD phase diagram at finite temperature and chemical potential \cite{Cai:2022omk,Cai:2024eqa,He:2013qq,Cai:2012xh,DeWolfe:2011ts}, for more related holographic studies, see~\cite{Zhao:2022uxc,Li:2023mpv,Zhao:2023gur,Liu:2023pbt,Chen:2024ckb,Fu:2024nmw,Toniato:2025gts,Jokela:2024xgz,Zhu:2025gxo,Lilani:2025wnd,Shen:2025yrn,Zeng:2025tcz,Jarvinen:2025mgj,Liu:2024efy,Arefeva:2024xmg,Arefeva:2024vom,Li:2025ugv,Deng:2026aht}. A well-known limitation of conventional EMD models, however, is that they belong to the class of two-derivative Einstein theories and therefore generically yield the universal result \(\eta/s = 1/(4\pi)\), which cannot account for the nontrivial temperature dependence of the shear viscosity suggested by phenomenology. To overcome this limitation, higher-derivative corrections, such as Gauss--Bonnet terms or curvature-squared invariants of the form \(R_{\mu\nu\rho\sigma}R^{\mu\nu\rho\sigma}\), have been introduced~\cite{Cai:2008ph1,Myers:2009ij,Cremonini:2009sy,Li:2014dsa,Buchel:2024umq}. By allowing the corresponding higher-derivative couplings to depend on the dilaton, such extensions can generate a temperature-dependent \(\eta/s\), and in some cases a non-monotonic behavior compatible with Bayesian constraints. However, reproducing the full phenomenologically favored structure often requires parameter choices that lie outside the perturbative regime of control, indicating that finite-coupling effects may play an essential role in QGP transport. Moreover, in other constructions~\cite{Buchel:2023fst,Apostolidis:2025gnn}, it is possible to obtain a temperature-dependent or non-monotonic \(\eta/s\), but generally at the price of losing simultaneous consistency with equation-of-state constraints. In this work, a Gauss--Bonnet extension of the EMD model~\cite{Cai:2022omk} is constructed, with the higher-derivative term retained non-perturbatively in the background dynamics. The formalism for the shear and bulk viscosities in Gauss--Bonnet--scalar theory was derived in~\cite{Tong:2025rxz}. With the model parameters fixed by lattice-QCD thermodynamics, the phase structure in the \(T\)-\(\mu_B\) plane is obtained, and the temperature dependence of \(\eta/s\) and \(\zeta/s\) is studied. The resulting holographic predictions are then confronted with phenomenological Bayesian constraints from heavy-ion collision analyses.

The remainder of this paper is organized as follows. Section~\ref{HQCD model} introduces the improved holographic QCD model and presents the gravitational setup, including the bulk action, the background ansatz, and the choice of model parameters. Sections~\ref{secHconstant} and~\ref{Hnonconstant} study the thermodynamics at finite temperature and baryon chemical potential, analyze observables such as the entropy density and the speed of sound, and determine the resulting phase structure and the location of the CEP. The holographic computation of the shear and bulk viscosities is also carried out in these sections, together with comparisons to phenomenological Bayesian constraints. Section~\ref{conclusion} summarizes the main results and discusses their implications and possible extensions. Technical details are collected in appendices~\ref{app:A}--\ref{app:viscosity}.

\section{Holographic QCD Model}
\label{HQCD model}

A longstanding challenge in holographic modeling is to reproduce both the equation of state and the transport coefficient \(\eta/s\) within a single framework. Two-derivative EMD models can achieve quantitative agreement with lattice-QCD results for the equation of state, but they generically give the universal value \(\eta/s = 1/(4\pi)\)~\cite{Cai:2022omk,Grefa:2022sav}. This value is not sufficient to describe the temperature-dependent behavior suggested by Bayesian analyses; see Fig.~\ref{fig:etaovers_1} below. In particular, the shear viscosity of the QGP is expected to decrease toward the crossover region and increase again away from it. Conversely, other holographic constructions~\cite{Li:2014dsa,Apostolidis:2025gnn} can generate a temperature-dependent \(\eta/s\) compatible with Bayesian constraints, but they do not simultaneously provide a lattice-constrained equation of state. This tension motivates a holographic framework in which thermodynamic and transport constraints are implemented consistently.

\subsection{Holographic Model}
The EMD framework is extended by incorporating higher-derivative corrections. The five-dimensional bulk action is
\begin{equation}
\label{five-dimensional bulk action}
S=\frac{1}{2\kappa_N^2}\int d^5x\,\sqrt{-g}\left[
R-\frac{1}{2}\nabla_\mu\phi\nabla^\mu\phi
-V(\phi)-\frac{Z(\phi)}{4}F_{\mu\nu}F^{\mu\nu}
+\alpha H(\phi)R_{\rm GB}^2
\right],
\end{equation}
where \(g\) is the determinant of the bulk metric \(g_{\mu\nu}\), \(\kappa_N^2\) denotes the five-dimensional Newton constant, and \(\phi\) is the dilaton field. In the holographic description, the dilaton encodes the nonconformal dynamics of QCD and represents the running of the effective coupling. The Maxwell field \(A_{\mu}\) is introduced to model finite baryon density. The functions \(V(\phi)\) and \(Z(\phi)\) specify the dilaton potential and the coupling between the Maxwell sector and the scalar background, respectively.

A central ingredient of the model is the Gauss--Bonnet term,
\begin{equation}
R_{\rm GB}^2
=
R_{\mu\nu\rho\sigma}R^{\mu\nu\rho\sigma}
-4R_{\mu\nu}R^{\mu\nu}
+R^2,
\end{equation}
which introduces higher-derivative corrections to the dynamics of gravity. In the present setup, this term generates a nontrivial temperature dependence of the shear viscosity to entropy density ratio \(\eta/s\), which is absent in conventional two-derivative EMD models. The coupling function \(H(\phi)\) is chosen to remain within the interval \([0,1]\). The parameter \(\alpha\) controls the overall strength of the Gauss--Bonnet contribution and regulates the impact of finite-coupling effects.

To study the thermodynamics of strongly coupled QCD matter holographically, a static and isotropic black-brane geometry is considered,
\begin{equation}
\begin{aligned}
ds^2 &= -f(r)e^{-\eta(r)}dt^2+\frac{dr^2}{f(r)}+r^2 (dx^2+dy^2+dz^2),\\
A_t&=A_t(r),\qquad \phi=\phi(r),
\end{aligned}
\label{eq:metric}
\end{equation}
where \(r\) is the holographic radial coordinate and the AdS boundary is located at \(r\to\infty\). The baryon chemical potential and baryon density are extracted from the near-boundary behavior of \(A_t\). The horizon is located at \(r=r_h\), where \(f(r_h)=0\). The Hawking temperature, identified with the temperature of the boundary theory, is fixed by the regularity of the Euclidean geometry and takes the form
\begin{equation}
T=\frac{1}{4\pi}f'(r_h)e^{-\eta(r_h)/2}.
\label{eqTands}
\end{equation}
Regularity of the Maxwell field further requires \(A_t(r_h)=0\).

In the presence of the Gauss--Bonnet term, the black-hole entropy is computed using the Wald formula~\cite{Wald:1993nt} rather than the Bekenstein--Hawking area law. For a general higher-derivative theory with the Lagrange density \(\mathcal{L}\), the Wald entropy is given by
\begin{equation}
S_{\rm Wald}
=
-2\pi \int_{\Sigma_h} d^3x\,\sqrt{h}\,
\frac{\partial \mathcal{L}}{\partial R_{\mu\nu\rho\sigma}}
\,\epsilon_{\mu\nu}\epsilon_{\rho\sigma},
\end{equation}
where \(\Sigma_h\) denotes the spatial section of the horizon, \(h\) is the induced metric on the horizon, and \(\epsilon_{\mu\nu}\) is the binormal to \(\Sigma_h\). In Einstein--Gauss--Bonnet gravity, the Wald entropy generally receives corrections from the intrinsic curvature of the horizon cross-section. For the planar black-brane geometry~\eqref{eq:metric}, however, the horizon is spanned by a flat spatial metric and its intrinsic curvature vanishes. As a result, the Gauss--Bonnet contribution to the Wald entropy is absent~\cite{Tong:2025rxz}, and the entropy reduces to the standard Bekenstein--Hawking form. The entropy density is therefore

\begin{equation}
s=\frac{2\pi}{\kappa_N^2}\,r_h^3\,.
\end{equation}

The explicit forms of the dilaton potential and the Maxwell coupling function are taken to be
\begin{equation}
\begin{aligned}
 H(\phi)=&1-\frac{h_1\ \phi^6}{h_2+\phi^6}\,,\\
V(\phi)=&-12\cosh(c_1\phi)+(6c_1^2-\tfrac{3}{2})\phi^2
+c_6\phi^3+c_2\phi^6+h_0\,,\\
Z(\phi)=&\frac{1}{1+c_3}\mathrm{sech}(c_4\phi^3)
+\frac{c_3}{1+c_3}e^{-c_5\phi}\,,
\end{aligned}
\label{eq:423}
\end{equation}
where \(c_{1-6}\) and \(h_{1-2}\) are phenomenological parameters determined by fitting lattice-QCD data, including the equation of state, the speed of sound, and other thermodynamic observables. The ansatz imposes that the first four derivatives of \(H(\phi)\) vanish at \(\phi=0\), ensuring that the higher-curvature coupling is sufficiently suppressed near the asymptotic AdS boundary. The Gauss--Bonnet coupling effectively shifts the cosmological constant and hence the effective AdS radius of the boundary geometry. To preserve the same asymptotic geometry, the constant \(h_0\) is introduced and fixed by
\begin{equation}
h_0=24\alpha\,.
\end{equation}
The functions \(V(\phi)\), \(Z(\phi)\), and \(H(\phi)\) determine the thermodynamic and transport properties of the model and are fixed by matching to lattice-QCD results at vanishing baryon chemical potential. The action~\eqref{five-dimensional bulk action} thus provides a unified framework for describing equilibrium thermodynamics and transport properties of strongly coupled QCD matter at finite temperature and baryon chemical potential.

\subsection{Thermodynamics}
Varying the action~\eqref{five-dimensional bulk action} with respect to the metric \(g_{\mu\nu}\), the dilaton field \(\phi\), and the Maxwell field \(A_\mu\), and substituting the black-brane ansatz~\eqref{eq:metric}, leads to a coupled system of ordinary differential equations for \(f(r)\), \(\eta(r)\), \(\phi(r)\), and \(A_t(r)\). The explicit equations are given in Appendix~\ref{app:A}.

The asymptotic expansion near the AdS boundary \(r\rightarrow\infty\) is given by
\begin{equation}
\begin{split}
\phi(r)&=\frac{\phi_s}{r}+\cdots+\frac{\phi_v}{r^3}+\cdots, \qquad A_t=\mu_B-\frac{\kappa_N^2 n_B}{r^2}+\cdots,\\
f(r)&=r^2(1+\cdots+\frac{f_v}{r^4}+\cdots),\qquad \eta(r)=0+\cdots,
\end{split}
\end{equation}
where \((\phi_s,\phi_v,\mu_B,n_B,f_v)\) are integration constants determined by solving the equations of motion. The coefficient \(\phi_s\) plays a special role as the source for the operator that explicitly breaks conformal invariance, thereby setting the characteristic energy scale of the model. In the numerical implementation, \(\phi_s\) is used to fix the overall scale for different parameter choices. The remaining UV coefficients are related to thermodynamic quantities of the dual field theory through the standard holographic dictionary; see Appendix~\ref{app:B} for details. In particular, \(\mu_B\) and \(n_B\) correspond to the baryon chemical potential and baryon number density, respectively. The energy density \(\epsilon\), pressure \(P\), and trace anomaly are
\begin{equation}
\begin{aligned}
\epsilon
=&\frac{1}{2\kappa_{N}^2}\Big(
\frac{\alpha \phi _s^4}{12 (1-4 \alpha )^2}
+\frac{\phi _s^4}{48 (1-4 \alpha )^2}
+b \phi _s^4
+\frac{27}{2} c_6^2 \phi _s^4
+3 (4 \alpha -1) f_v
+\phi _s \phi _v
\Big),\\
P
=&\frac{1}{2\kappa_{N}^2}\Big(
(4 \alpha -1) f_v
+\phi_s \phi _v
+3 c_6^2 \phi _s^4
-b \phi _s^4
-\frac{\phi _s^4 \left(28 \alpha +24 (1-4 \alpha )^2 c_1^4-9\right)}{144 (1-4 \alpha )^2}
\Big),\\
\epsilon-3P
=&\frac{1}{2\kappa_{N}^2}\Big(
\frac{2 \alpha \phi_s^4}{3 (1-4 \alpha )^2}
-\frac{\phi_s^4}{6 (1-4 \alpha )^2}
+4 b \phi_s^4
+\frac{1}{2} c_1^4 \phi_s^4
+\frac{45}{2} c_6^2 \phi_s^4
-2 \phi_s \phi_v
\Big).
\end{aligned}
\label{eq:Tands}
\end{equation}
The free energy density satisfies \(\Omega=-P=\epsilon-Ts-\mu_B n_B\). Other thermodynamic observables, including the squared speed of sound \(c_s^2\) and the specific heat \(C_V\), can be obtained from these quantities and are useful for characterizing the QCD transition. The second-order baryon susceptibility,
\begin{equation}
  \chi_2^B  =\frac{1}{T^2}  \left(\frac{\partial n_{B}}{\partial\mu_{B}} \right)_{T}\,,
\end{equation}
is particularly important for constraining the gauge kinetic function \(Z(\phi)\).

\subsection{Shear and bulk viscosities}
\label{Shear and bulk viscosities}

This subsection summarizes the holographic computation of the shear and bulk viscosities. Technical details are provided in Appendix~\ref{app:viscosity}. Both quantities are real-time transport coefficients and can be extracted from Kubo formulas involving retarded Green's functions of the energy-momentum tensor.

For the shear viscosity, the tensor fluctuation in the \(xy\) channel is considered. The standard Kubo formula is
\begin{equation}
\eta=-\lim_{\omega\to0}\frac{1}{\omega}\,
\mathrm{Im}\,G^{R}_{xy,xy}(\omega,\vec{k}=0)\,.
\label{eq:etaKubo_main}
\end{equation}
Following the prescription of~\cite{Myers:2009ij}, the low-frequency computation reduces to the canonical momentum associated with the transverse graviton mode. Applied to the action~\eqref{five-dimensional bulk action}, this method gives the shear viscosity to entropy density ratio
\begin{equation}
\frac{{\eta }}{s} = \frac{1}{4\pi }\Big[1- \frac{2 \alpha}{r_h} f'(r_h) \big(\phi '(r_h) \partial_{\phi} H(\phi (r_h))+H(\phi (r_h))\big)\Big].
\label{eq:etasfinal_main}
\end{equation}
This expression shows explicitly how the Gauss--Bonnet sector modifies the universal KSS value \({{\eta }}/{s}=1/(4\pi)\). The deviation is governed by the higher-derivative coupling \(\alpha H(\phi)\) and by the horizon data of the background geometry. In the limit \(\alpha\to0\), the standard result \(\eta/s=1/(4\pi)\) is recovered.


To obtain the bulk viscosity we consider SO(3) invariant perturbations of the metric and dilaton. The corresponding Kubo formula is
\begin{equation}
\zeta
=
-\frac{4}{9}\lim_{\omega\to0}\frac{1}{\omega}\,
\mathrm{Im}\,G_R(\omega),
\label{eq:zetaKubo_main}
\end{equation}
where \(G_R(\omega)\) denotes the retarded Green's function in the bulk channel. Following~\cite{Buchel:2023fst, Tong:2025rxz}, a gauge-invariant scalar fluctuation is introduced, and its linearized equation is solved on the background geometry~\eqref{eq:metric}. In the small-frequency expansion, the imaginary part of the conserved radial flux determines the bulk viscosity, leading to
\begin{equation}
\frac{\zeta}{s}
=
\frac{1}{4\pi}\cdot\frac{4}{9}\,(z_0^h)^2,
\label{eq:zetasfinal_main}
\end{equation}
where \(z_0^h\) is the zero-frequency value of the gauge-invariant scalar fluctuation evaluated at the horizon. The derivation of the fluctuation equations and the numerical implementation are presented in Appendix~\ref{bulkviscosity}.

A technical point concerns the treatment of the bulk-channel fluctuation equation. In this channel, the equation is expanded perturbatively in the Gauss--Bonnet coupling \(\alpha\), and only the leading two terms, \emph{i.e.}, \(O(\alpha^0)\) and \(O(\alpha^1)\), are retained. Higher-order corrections are neglected. Therefore, the numerical results for \(\zeta/s\) presented below should be understood within this \(O(\alpha)\) approximation for the bulk channel. By contrast, the thermodynamic background is obtained from the full equations of motion. Since the fitted value of \(\alpha\) is small, the perturbative treatment in the bulk channel is expected to be self-consistent.

\section{Application: the case of constant \(H(\phi)\)}
\label{secHconstant}

The simplest benchmark realization corresponds to a constant Gauss--Bonnet coupling function \(H(\phi)\). Without loss of generality, this case is specified by
\begin{equation}\label{Hconstant}
H(\phi)=1\,.
\end{equation}
This choice isolates the effect of the Gauss--Bonnet coupling \(\alpha\), without introducing additional scalar dependence through \(H(\phi)\). It therefore provides a useful baseline for assessing how higher-curvature corrections affect the thermodynamics and transport properties of the holographic plasma.

\subsection{Thermodynamics and phase diagram}
\label{Thermodynamics and phase diagram}
The thermodynamic sector is first analyzed for the benchmark choice~\eqref{Hconstant}. The model parameters are fixed by a global fit to lattice-QCD data at \(\mu_B=0\), requiring the resulting equation of state and other equilibrium observables to remain consistent with available lattice constraints. Simultaneously, the Gauss--Bonnet coupling is chosen such that \(\eta/s\) lies within the \(90\%\) credible interval inferred from the JETSCAPE Bayesian analysis. This procedure yields a parameter set that is consistent with both thermodynamic constraints and phenomenological transport bounds.

\begin{figure}[htbp]
\centering
\includegraphics[width=0.8\textwidth]{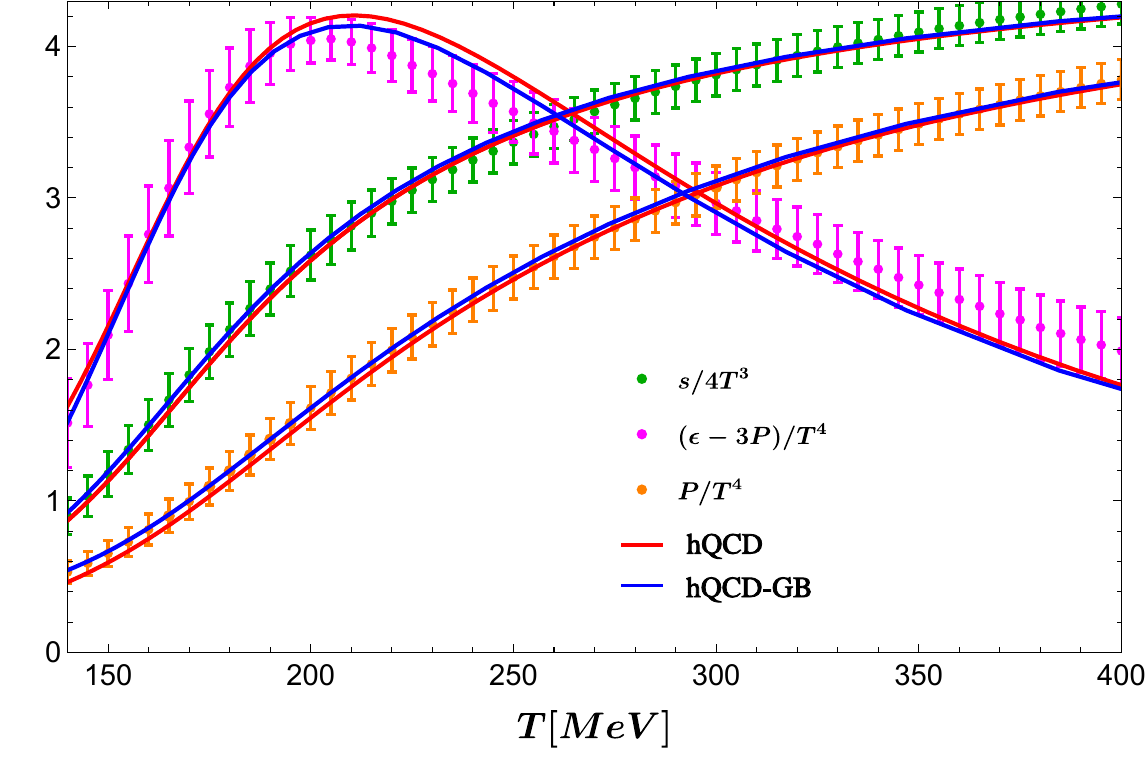}
\caption{\label{fig:eos}
Equation of state at \(\mu_B=0\). The blue solid curve denotes the result obtained from the present model with \(H(\phi)=1\), while the red solid curve denotes the EMD model~\cite{Cai:2022omk}. The error bars correspond to lattice-QCD data~\cite{HotQCD:2014kol,Borsanyi:2021sxv}.
}
\end{figure}

\begin{figure}[htbp]
\centering
\includegraphics[width=0.8\textwidth]{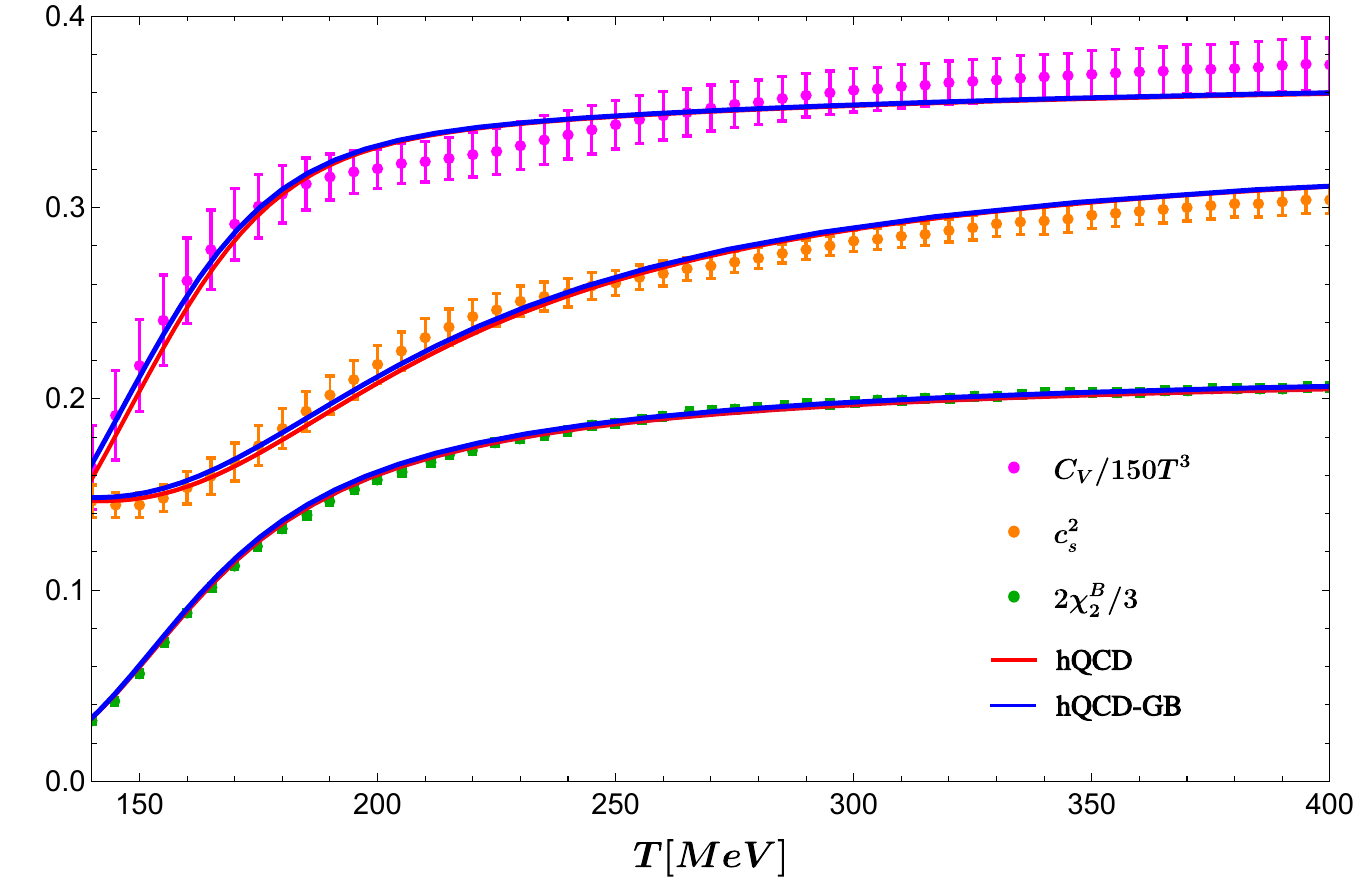}
\caption{\label{fig:cs2}
Temperature dependence of the specific heat \(C_V\), squared speed of sound \(c^2_s\), and baryon susceptibility \(\chi^B_2\) at \(\mu_B=0\). The blue and red solid curves correspond to the EMD model~\cite{Cai:2022omk} and the present model with \(H(\phi)=1\), respectively. The error bars correspond to lattice-QCD data~\cite{HotQCD:2014kol,Borsanyi:2021sxv}.
}
\end{figure}

The resulting parameters are listed in Table~\ref{tab:parameters1} for \(\alpha=-0.006\), and the equation of state at \(\mu_B=0\) is shown in Fig.~\ref{fig:eos}. The pressure, entropy density, and trace anomaly obtained from the Gauss--Bonnet-corrected model remain in close agreement with the corresponding lattice data over the temperature range of interest. Fig.~\ref{fig:cs2} shows additional thermodynamic observables, including the squared speed of sound, specific heat, and baryon susceptibility. These results indicate that the constant-coupling Gauss--Bonnet term preserves the successful thermodynamic matching of the EMD construction~\cite{Cai:2022omk}. In particular, the trace anomaly is better reproduced than in the corresponding EMD model.

\begin{table}[h]
\centering
\begin{tabular}{|c|c|c|c|c|c|c|c|c|}
\hline
\(c_{1}\) & \(c_{2}\) & \(c_{3}\) & \(c_{4}\) & \(c_{5}\) & \(c_{6}\) & \(b\) & \(\phi_{s}~[\text{MeV}]\) & \(\kappa_{N}^{2}\) \\
\hline
0.7074 & 0.0041 & 1.95 & 0.09 & 30 & 0 & -0.2744 & 1065 & \(2\pi(1.62)\) \\
\hline
\end{tabular}
\caption{\label{tab:parameters1}
Model parameters for the benchmark case \(H(\phi)=1\) with \(\alpha=-0.006\).
}
\end{table}

\begin{figure}[htbp]
\centering
\includegraphics[width=1\textwidth]{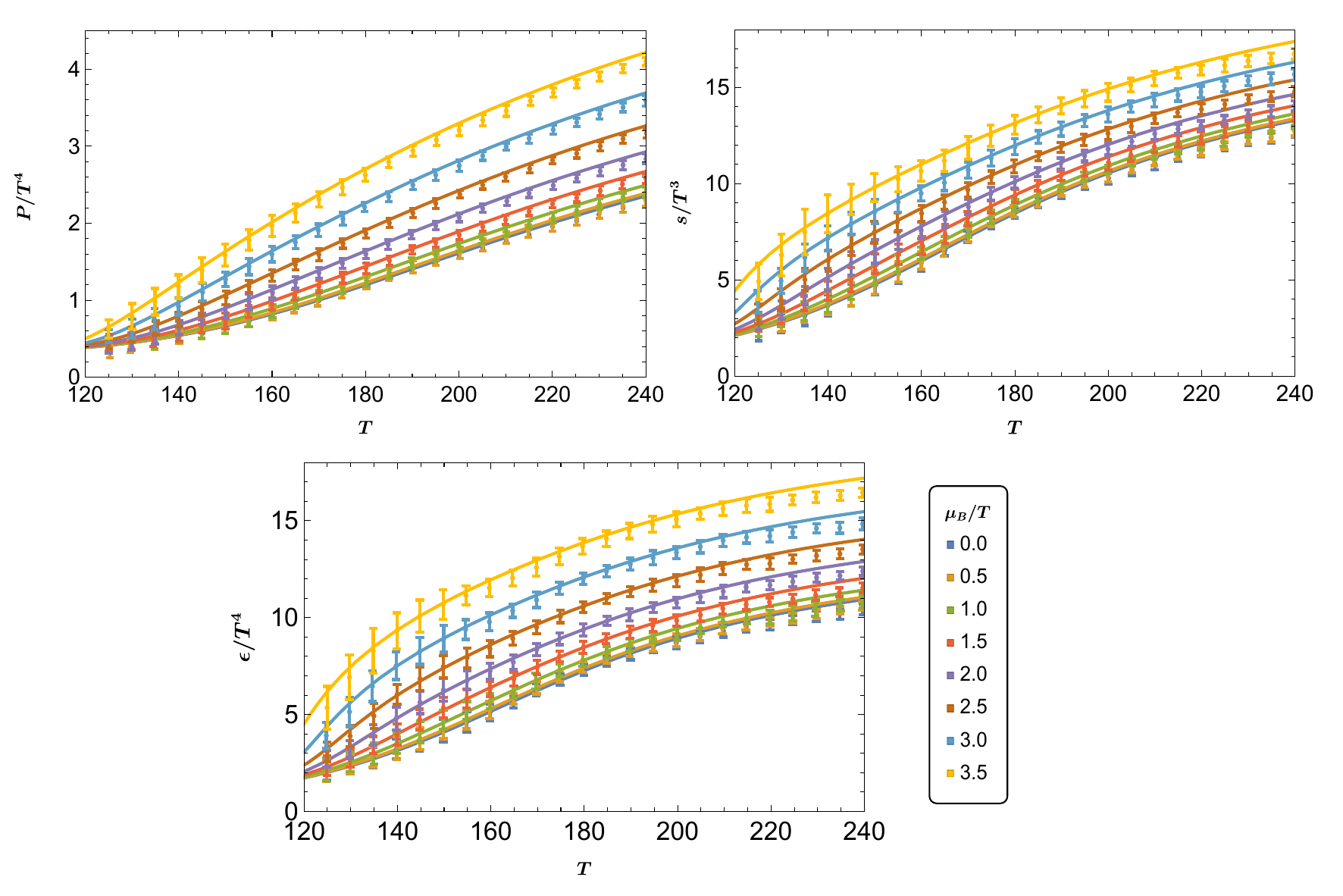}
\caption{\label{fig:fixmu}
Thermodynamic observables at finite baryon chemical potential: pressure \(P\), entropy density \(s\), and energy density \(\epsilon\). The solid curves denote the results from the present model and are compared with lattice-QCD data~\cite{Borsanyi:2021sxv}. In each panel, the curves span the range \(0\le \mu_B/T\le 3.5\), with lower and upper curves corresponding to smaller and larger values of \(\mu_B/T\), respectively.
}
\end{figure}

After the parameters are fixed at \(\mu_B=0\), the analysis is extended to finite baryon chemical potential. Using the same parameter set, the thermodynamic observables are computed as functions of temperature for several values of \(\mu_B/T\) in the range \(0\le \mu_B/T\le 3.5\). The results are shown in Fig.~\ref{fig:fixmu}, where the holographic predictions for the pressure, entropy density, and energy density are compared with available lattice-QCD data~\cite{HotQCD:2014kol,Borsanyi:2021sxv}. The overall agreement remains qualitatively good throughout this range, indicating that the constant-coupling Gauss--Bonnet extension retains a reasonable description of the finite-density equation of state.

\begin{figure}[htbp]
\centering
\includegraphics[width=1\textwidth]{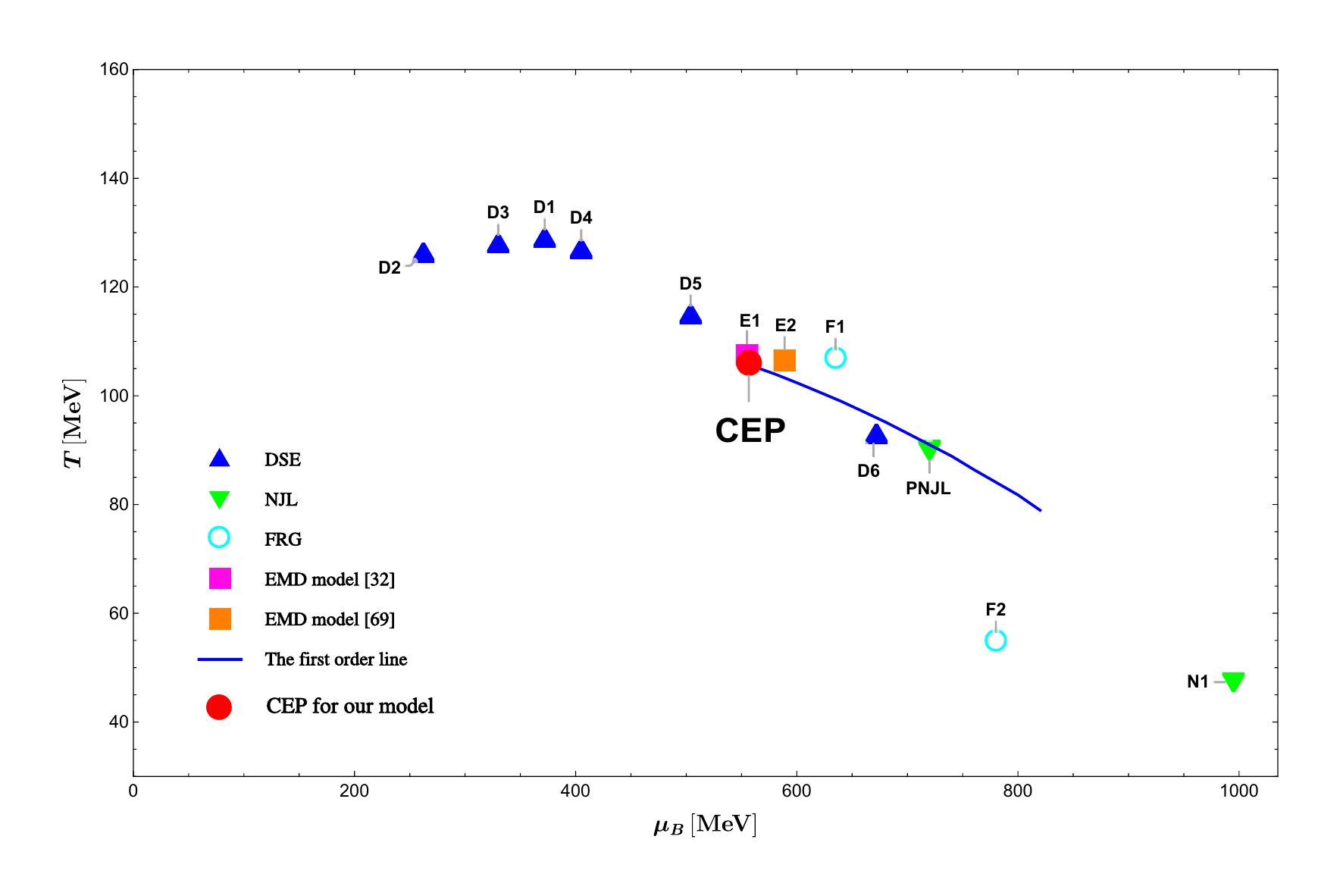}
\caption{\label{phase diagram}
QCD phase diagram predicted by the present model. The first-order phase transition line, shown as the solid blue line, is determined from the free energy. The CEP is located at \(T_{C}=105~\text{MeV}\) and \(\mu_{C}=556~\text{MeV}\), indicated by the bold red dot. The pink cube E1 denotes the CEP from the holographic EMD model~\cite{Cai:2022omk}, while the orange cube E2 denotes the CEP from~\cite{Hippert:2023bel}. CEP locations obtained from other approaches are also shown, including the Schwinger--Dyson equation (DSE) approach (D1--D7 from~\cite{Xin:2014ela,Gao:2016qkh,Qin:2010nq,Shi:2014zpa,Fischer:2014ata,Gao:2020qsj,Lu:2025cls}), the Nambu--Jona-Lasinio model (NJL), including the PNJL result from~\cite{Li:2018ygx} and the NJL1 result from~\cite{Asakawa:1989bq}, and the functional renormalization group (FRG) method (F1--F3 from~\cite{Fu:2019hdw,Zhang:2017icm,Fu:2023lcm}).
}
\end{figure}

Based on these thermodynamic results, the phase structure in the \(T\)-\(\mu_B\) plane is determined. The resulting phase diagram is shown in Fig.~\ref{phase diagram}. The CEP is located at
\begin{equation}
T_C=105~\text{MeV},
\qquad
\mu_C=556~\text{MeV},
\end{equation}
as indicated by the red point in Fig.~\ref{phase diagram}. Within the present model, no first-order phase transition appears for \(\mu_B/T<\mu_C/T_C\approx 5.295\). 

For comparison, Fig.~\ref{phase diagram} also shows CEP locations obtained from other approaches. The present result is close to those from Schwinger--Dyson equations~\cite{Fischer:2014ata,Gao:2020qsj}, the functional renormalization group~\cite{Fu:2019hdw}, and holographic EMD models~\cite{Cai:2022omk,Hippert:2023bel}. In particular, it is nearly identical to the CEP obtained in the EMD model~\cite{Cai:2022omk}, where \(T_C=105~\text{MeV}\) and \(\mu_C=555~\text{MeV}\). This proximity is consistent with the small value of the Gauss--Bonnet parameter used here, \(\alpha=-0.006\), for which the higher-derivative correction only mildly modifies the thermodynamic background.

\subsection{Shear viscosity}

\begin{figure}[htbp]
\centering
\includegraphics[width=0.8\textwidth]{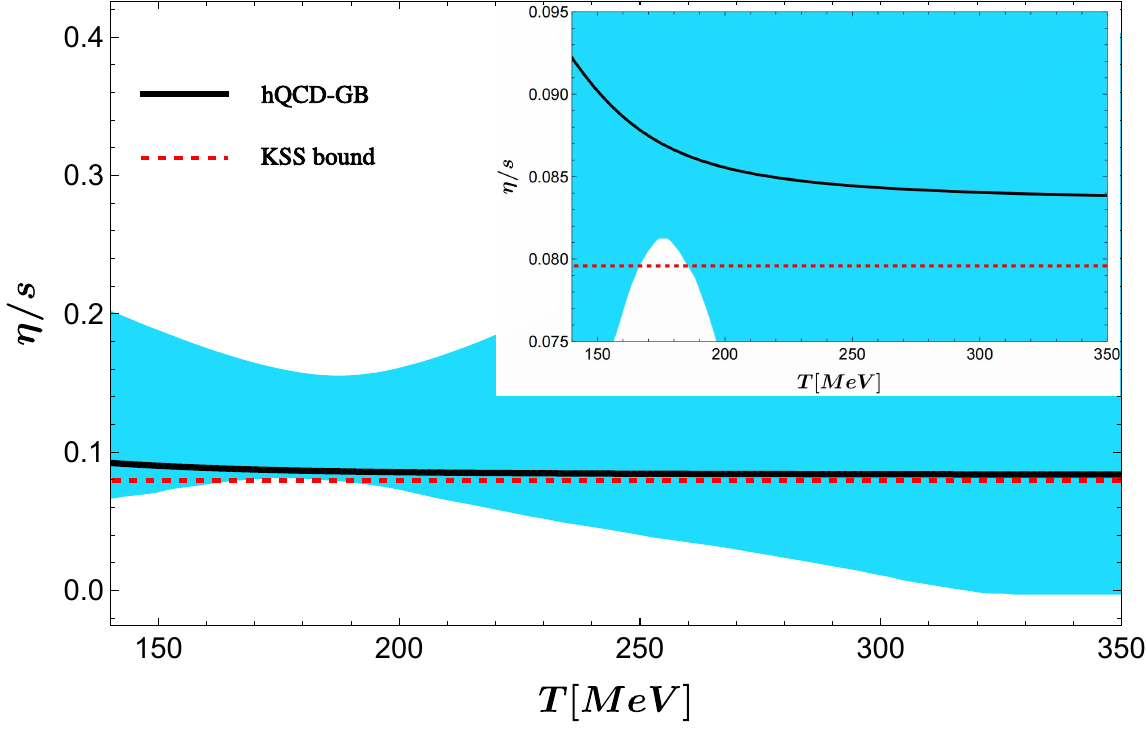}
\caption{\label{fig:etaovers_1}
Temperature dependence of the shear viscosity to entropy density ratio \(\eta/s\) in the Gauss--Bonnet-corrected model with \(\alpha=-0.006\) and \(H(\phi)=1\). The black solid curve denotes the holographic result, the red dashed line indicates the universal KSS value \(\eta/s=1/(4\pi)\), and the blue band represents the \(90\%\) credible interval from JETSCAPE~\cite{JETSCAPE:2020mzn}.
}
\end{figure}

After establishing that the constant-coupling model provides an adequate description of equilibrium thermodynamics, the shear viscosity is analyzed. In this setup, the higher-derivative contribution is encoded in the Gauss--Bonnet correction, which directly affects the shear channel. For \(H(\phi)=1\), the expression for the shear viscosity to entropy density ratio in~\eqref{eq:etasfinal_main} reduces to
\begin{equation}
\begin{aligned}
 \frac{\eta}{s} = \frac{1}{4\pi }\Big(1-\frac{2 \alpha  f'(r_h) }{r_h}\Big)\,.
\end{aligned}
    \label{eq:473h=1}
\end{equation}
Thus, even in the constant-coupling case, \(\eta/s\) is no longer universal but is determined by the horizon data of the black-brane geometry~\eqref{eq:metric}. The resulting temperature dependence is shown by the black solid curve in Fig.~\ref{fig:etaovers_1}. The Gauss-Bonnet correction shifts \(\eta/s\) away from the universal KSS value, indicated by the red dashed line, and generates a nontrivial thermal profile. For \(\alpha=-0.006\), \(\eta/s\) decreases monotonically with increasing temperature and remains broadly compatible with the phenomenological range inferred from the JETSCAPE Bayesian analysis~\cite{JETSCAPE:2020mzn}. However, this monotonic behavior does not reproduce the expected minimum of \(\eta/s\) near the QCD transition, as suggested in~\cite{Nakamura:2004sy, Csernai:2006zz}, where the transport behavior of QCD matter was compared with that of ordinary fluids near phase transitions. This limitation motivates the introduction of a scalar-dependent Gauss--Bonnet coupling \(H(\phi)\), which provides additional flexibility to model a minimum of \(\eta/s\) near the crossover region.

\subsection{Bulk viscosity}
\label{bulk viscosity h=1}

\begin{figure}[htbp]
\centering
\includegraphics[width=0.8\textwidth]{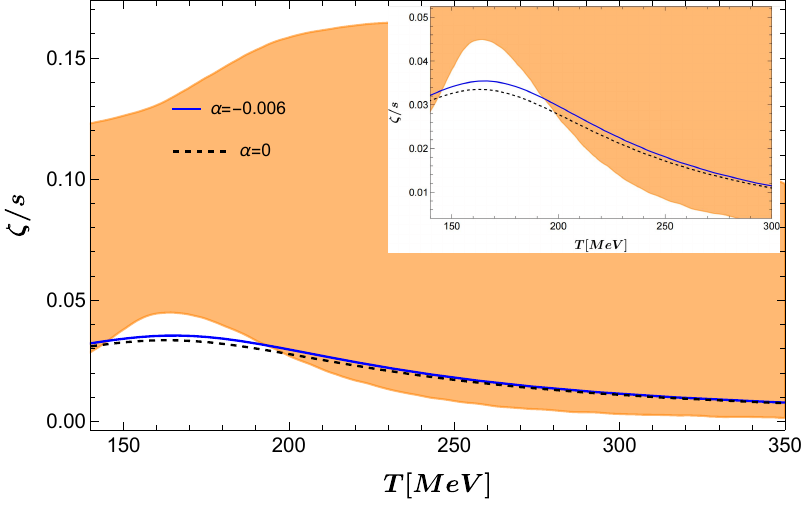}
\caption{\label{fig:zetaaovers_1}
Bulk viscosity to entropy density ratio \(\zeta/s\) as a function of temperature. The blue solid curve corresponds to the Gauss--Bonnet-corrected model with \(H(\phi)=1\) and \(\alpha=-0.006\), while the dashed curve denotes the case without higher-derivative corrections. The orange band represents the \(90\%\) credible interval from the JETSCAPE analysis~\cite{JETSCAPE:2020mzn}.
}
\end{figure}

The bulk viscosity is next analyzed in the constant-coupling setup. Unlike the shear viscosity, the bulk viscosity is directly sensitive to the nonconformal dynamics of the plasma and therefore provides a stringent test of the model near the crossover region. The ratio \(\zeta/s\) is computed using~\eqref{eq:zetasfinal_main}. As explained in Section~\ref{Shear and bulk viscosities}, the bulk-channel fluctuation equation is treated within the \(O(\alpha)\) approximation, retaining only the \(O(\alpha^0)\) and \(O(\alpha^1)\) terms in the expansion in the Gauss--Bonnet coupling. The numerical result for \(\alpha=-0.006\), obtained from~\eqref{eq:zetasfinal_main}, is shown as the blue solid curve in Fig.~\ref{fig:zetaaovers_1}. For comparison, the dashed curve denotes the case without higher-derivative corrections, \(\alpha=0\), using the same parameter set as in Table~\ref{tab:parameters1}. The orange band represents the \(90\%\) credible interval from the JETSCAPE analysis~\cite{JETSCAPE:2020mzn}.

The results show two main features. First, both curves are broadly compatible with the phenomenological range reported by JETSCAPE, indicating that the model captures the correct order of magnitude of the bulk viscosity in the relevant temperature window. Second, the inclusion of the Gauss--Bonnet term improves the agreement with the phenomenological constraint: the \(\alpha=-0.006\) result lies systematically closer to the JETSCAPE band than the corresponding two-derivative result with \(\alpha=0\). This suggests that higher-curvature corrections can have a quantitative impact on the bulk channel, in addition to their direct effect on the shear channel.

Several limitations should also be noted. A more negative value of \(\alpha\) could further improve the fit for \(\zeta/s\), but such a choice may compromise other aspects of the model. Thus, although the constant-coupling Gauss--Bonnet correction improves the bulk channel, it does not provide a fully adequate global description. Moreover, while \(\eta/s\) remains within the phenomenologically allowed range from the JETSCAPE analysis, its temperature dependence is still too simple to reproduce the non-monotonic structure expected for QCD matter, in particular, the possible minimum of \(\eta/s\) near the crossover region.

The constant-coupling ansatz should therefore be regarded as a useful benchmark rather than a complete quantitative description of the viscosities. The case \(H(\phi)=1\) nevertheless shows that higher-curvature corrections can improve both shear and bulk viscosities while preserving a reasonable fit to lattice-QCD thermodynamics. Residual discrepancies remain, especially in the detailed temperature dependence of the transport coefficients. This motivates a more general setup in which the Gauss--Bonnet coupling depends nontrivially on the scalar field.

\section{Application: the case of non-constant \(H(\phi)\)}\label{Hnonconstant}
The constant-coupling analysis shows that Gauss-Bonnet corrections can break the universality of \(\eta/s\) and bring its magnitude closer to phenomenological constraints. However, the resulting temperature dependence remains monotonic and therefore does not capture the expected minimum of \(\eta/s\) near the QCD transition. A natural extension is to allow the Gauss--Bonnet coupling to depend on the dilaton, so that the higher-derivative correction becomes sensitive to the scalar background that encodes the plasma's nonconformal dynamics. This section considers a non-constant Gauss--Bonnet coupling \(H(\phi)\). Compared with the benchmark case \(H(\phi)=1\), the additional parameters \(h_1\) and \(h_2\) in~\eqref{eq:423} enlarge the model space and make a complete parameter search more involved. Rather than performing a global optimization over all parameter choices or functional forms, a simple ansatz is adopted to assess the impact of a scalar-dependent Gauss--Bonnet coupling on thermodynamics and transport coefficients.
Specifically,
\begin{equation}
\label{h1h2}
h_1=h_2=0.5\,,
\end{equation}
is chosen as a representative parameterization, rather than a unique or optimal choice. A systematic, data-driven determination of \(H(\phi)\) is left for future work~\cite{Cai:2024eqa}. The resulting holographic model is analyzed following the same procedure as in the constant-coupling case. The thermodynamic properties are examined first, providing the background for the subsequent computation of the shear and bulk viscosities.

\subsection{Thermodynamics}

In the non-constant \(H(\phi)\) setup, the model parameters are again determined by matching lattice-QCD data for the equation of state and related observables at vanishing baryon chemical potential. The resulting parameter set is listed in Table~\ref{tab:parameters2}. Compared with the constant-coupling case in Table~\ref{tab:parameters1}, the fitted parameters change only moderately, indicating that the introduction of a nontrivial \(H(\phi)\) preserves the thermodynamic consistency of the model.

\begin{figure}[htbp]
\centering
\includegraphics[width=0.8\textwidth]{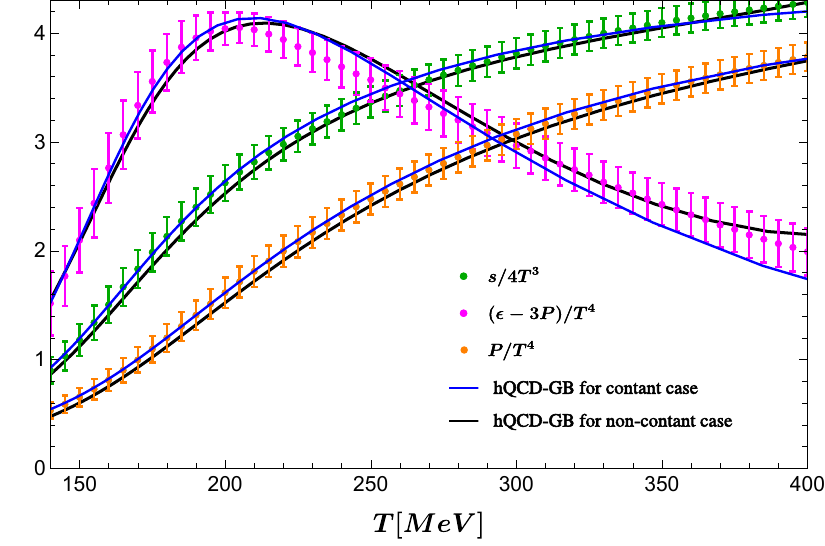}
\caption{\label{fig:eos_h_nonconstant}
Equation of state at vanishing baryon chemical potential for the non-constant-\(H(\phi)\) model. The black solid curve denotes the Gauss--Bonnet-corrected result with non-constant \(H(\phi)\), while the blue solid curve corresponds to the Gauss--Bonnet-corrected model with \(H(\phi)=1\). The error bars correspond to lattice-QCD data~\cite{HotQCD:2014kol, Borsanyi:2021sxv}.
}
\end{figure}

\begin{figure}[htbp]
\centering
\includegraphics[width=0.8\textwidth]{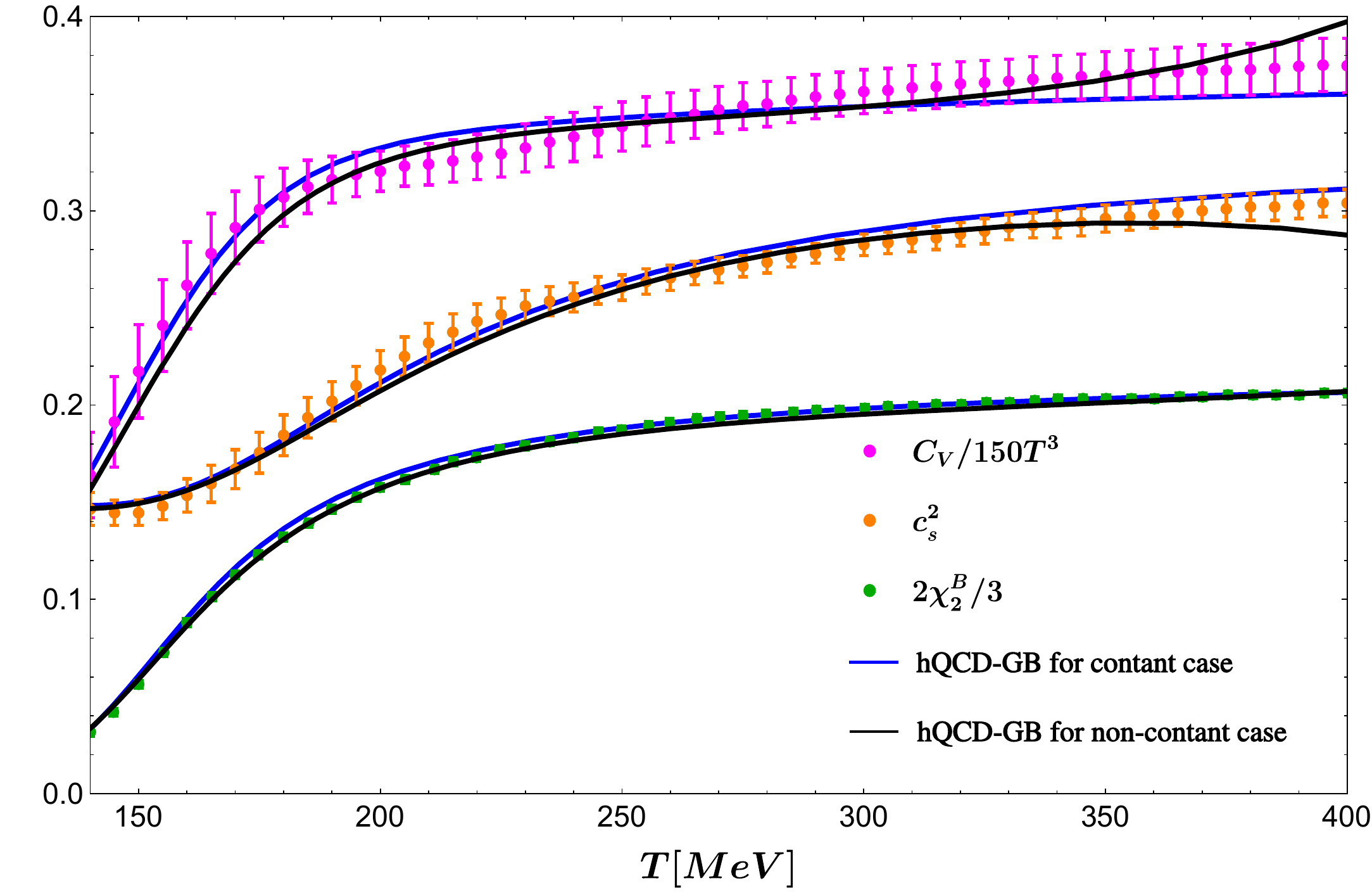}
\caption{\label{fig:cscv_h_nonconstant}
Thermodynamic observables at \(\mu_B=0\) for the non-constant-\(H(\phi)\) model. Shown are the squared speed of sound, the specific heat, and the baryon number susceptibility in the Gauss--Bonnet-corrected model with scalar-dependent coupling. The black and blue solid curves denote the non-constant-\(H(\phi)\) model and the Gauss--Bonnet-corrected model with \(H(\phi)=1\), respectively. The error bars correspond to lattice-QCD data~\cite{HotQCD:2014kol,Borsanyi:2021sxv}.
}
\end{figure}

The corresponding thermodynamic results at \(\mu_B=0\) are shown in Figs.~\ref{fig:eos_h_nonconstant} and~\ref{fig:cscv_h_nonconstant}. Fig.~\ref{fig:eos_h_nonconstant} displays the equation of state, while Fig.~\ref{fig:cscv_h_nonconstant} shows the squared speed of sound, specific heat, and baryon susceptibility. The black solid curves denote the predictions of the Gauss--Bonnet-corrected model with non-constant \(H(\phi)\), while the blue solid curves correspond to the constant-coupling case \(H(\phi)=1\) discussed in Section~\ref{Thermodynamics and phase diagram}. The present parameterization reproduces the lattice-QCD constraints well, showing that the scalar dependence introduced through \(H(\phi)\) can be incorporated without spoiling the agreement with equilibrium thermodynamics. At high temperatures, however, the model becomes more sensitive to temperature variations. For \(T>380\) MeV, both the speed of sound and the specific heat begin to deviate from the lattice-QCD constraints. Therefore, the subsequent analysis of transport coefficients is restricted to \(T<350\) MeV, where the model remains consistent with lattice-QCD thermodynamics.

\begin{table}[h]
\centering
\begin{tabular}{|c|c|c|c|c|c|c|c|c|}
\hline
\(c_{1}\) & \(c_{2}\) & \(c_{3}\) & \(c_{4}\) & \(c_{5}\) & \(c_{6}\) & \(b\) & \(\phi_{s}~[\text{MeV}]\) & \(\kappa_{N}^{2}\) \\
\hline
0.733 & 0.0046 & 2.075 & 0.07 & 30 & 0.10957 & -0.2744 & 985 & \(2\pi(1.49)\) \\
\hline
\end{tabular}
\caption{\label{tab:parameters2}
Model parameters for the non-constant-coupling case with \(h_1=h_2=0.5\) in \(H(\phi)\) and \(\alpha=-0.008\).
}
\end{table}

\begin{figure}[htbp]
\centering
\includegraphics[width=1.0\textwidth]{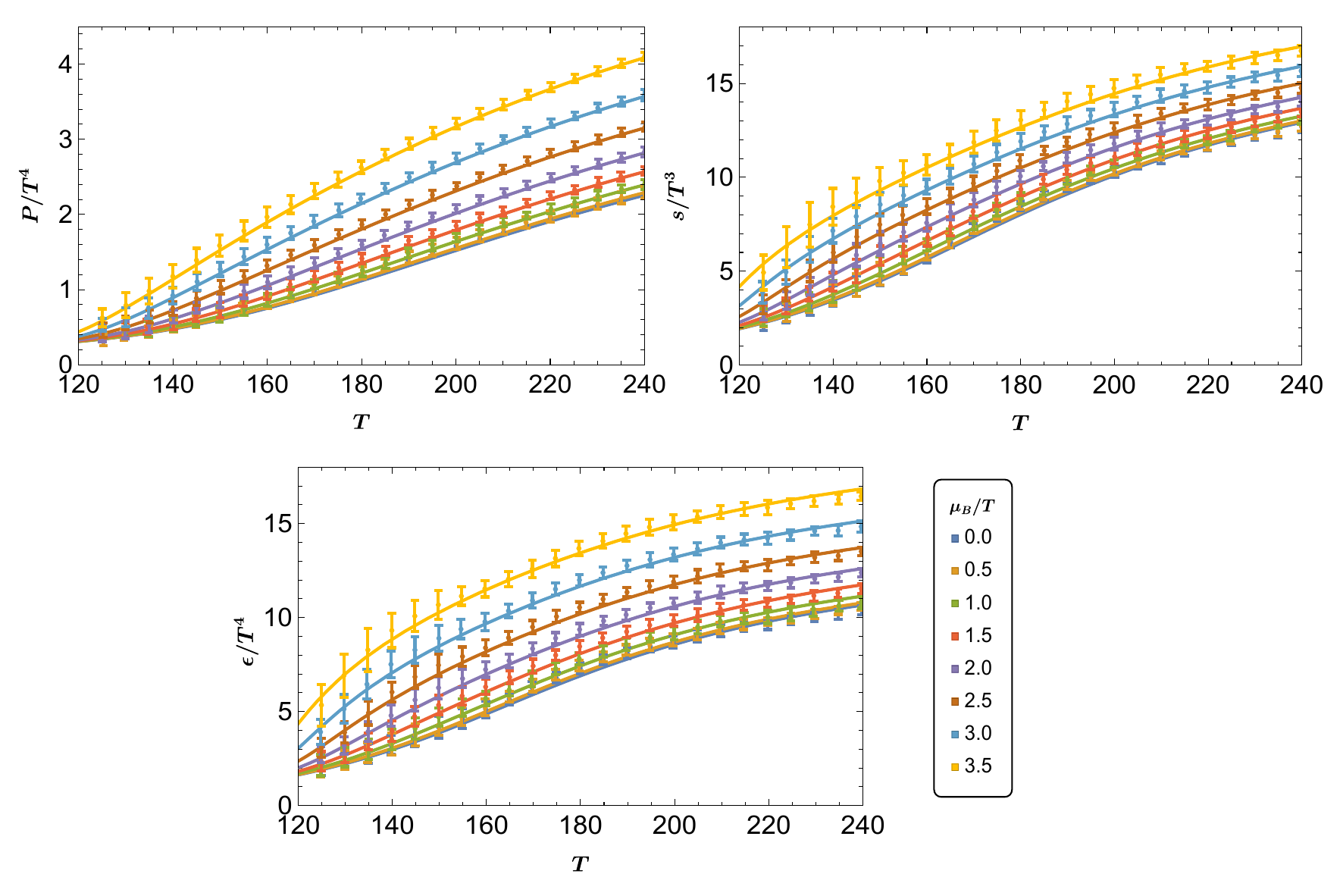}
\caption{\label{fig:fixmu2}
Thermodynamic observables at finite baryon chemical potential for the non-constant-\(H(\phi)\) model with \(h_1=h_2=0.5\): pressure \(P\), entropy density \(s\), and energy density \(\epsilon\). In each panel, the curves span the range \(0\le \mu_B/T\le 3.5\), with lower and upper curves corresponding to smaller and larger values of \(\mu_B/T\), respectively. The error bars are taken from lattice-QCD data~\cite{Borsanyi:2021sxv}.
}
\end{figure}

As a nontrivial consistency check, the analysis is extended to a finite baryon chemical potential. Fig.~\ref{fig:fixmu2} shows the resulting equation of state at finite \(\mu_B\), where the holographic predictions are compared with available lattice-QCD data. The solid curves correspond to the non-constant-\(H(\phi)\) model. The comparison indicates that the scalar-dependent Gauss--Bonnet coupling preserves a good overall description of finite-density thermodynamics, with deviations from lattice data remaining within the expected uncertainties. Thus, the improvement in the transport sector discussed below is not obtained at the expense of the equation of state. Instead, the model maintains a reliable thermodynamic baseline at both zero and finite baryon chemical potential.

\subsection{Shear viscosity}

For the parameter choice~\eqref{h1h2}, the shear viscosity to entropy density ratio follows from~\eqref{eq:etasfinal_main}:
\begin{equation}
\frac{\eta}{s}=\frac{1}{4\pi}\left(1-\frac{2\alpha f'(r_h)}{r_h}\frac{1 + 3 \phi(r_h)^6 + 2 \phi(r_h)^{12}
- 6 r_h \phi(r_h)^5 \phi'(r_h)
}{
\left(1 + 2 \phi(r_h)^6\right)^2
}\right)\,,
\label{shear viscosityhphi}
\end{equation}
This expression shows that \(\eta/s\) is controlled not only by the horizon data of the background geometry but also by the scalar dependence of \(H(\phi)\). This additional dependence allows for a richer temperature profile than in the constant-coupling case. In the limit of constant \(H(\phi)\), the expression reduces to~\eqref{eq:473h=1}.

\begin{figure}[htbp]
\centering
\includegraphics[width=0.8\textwidth]{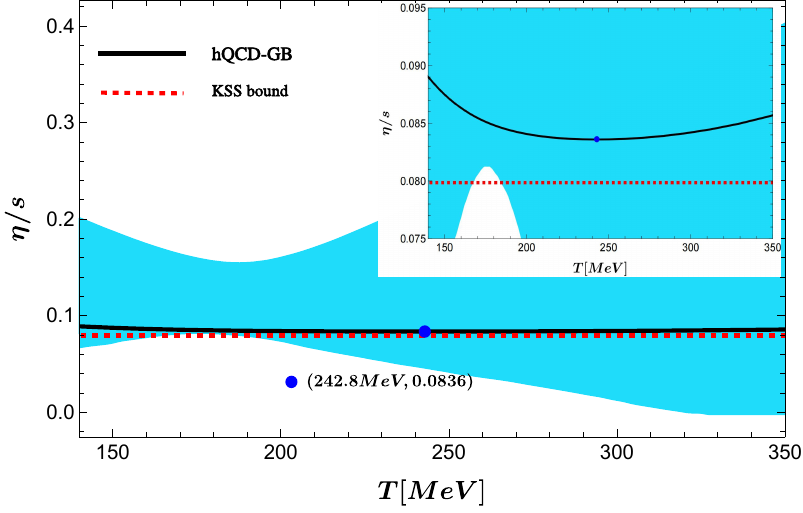}
\caption{\label{fig:etaovers_h_nonconstant}
Temperature dependence of the shear viscosity to entropy density ratio \(\eta/s\) for the model with nonconstant \(H(\phi)\). The blue dot represents the minimum value of $\eta/s$. The black solid curve shows the holographic result obtained in the present model, the red dashed line denotes the universal KSS bound \(\eta/s=1/(4\pi)\), and the blue band is  $90\%$ credible intervals from  JETSCAPE~\cite{JETSCAPE:2020mzn}.
}
\end{figure}

Using~\eqref{shear viscosityhphi} and the parameter set in Table~\ref{tab:parameters2}, \(\eta/s\) is computed numerically. The result is shown in Fig.~\ref{fig:etaovers_h_nonconstant}. The black solid curve denotes the holographic result, the red dashed line marks the universal Einstein-gravity value \(\eta/s = 1/(4\pi)\), and the blue band represents the \(90\%\) credible interval from JETSCAPE~\cite{JETSCAPE:2020mzn}. The ratio \(\eta/s\) exhibits a weak but visible nonmonotonic temperature dependence: it first decreases with increasing temperature, reaches a minimum
\begin{equation}
\left.\frac{\eta}{s}\right|_{\rm min}=0.0836
\qquad \text{at} \qquad
T=248.8~\mathrm{MeV},
\end{equation}
and then increases at higher temperatures.

This behavior differs qualitatively from the constant case \(H(\phi)=1\)  shown in Fig.~\ref{fig:etaovers_1}. The scalar-dependent Gauss--Bonnet coupling therefore provides the flexibility needed to generate a more realistic temperature dependence of the shear viscosity. The increase of \(\eta/s\) away from the minimum, both toward lower and higher temperatures, is consistent with the expectation that momentum transport becomes more dissipative outside the crossover region.

The resulting \(\eta/s\) curve remains within the \(90\%\) credible interval obtained from the Bayesian analysis of~\cite{JETSCAPE:2020mzn}. This indicates that the scalar-dependent coupling improves the qualitative thermal behavior of \(\eta/s\) while maintaining quantitative consistency with current phenomenological constraints. Since the nonmonotonic behavior is obtained without spoiling the thermodynamic agreement discussed above, the non-constant \(H(\phi)\) model provides a useful starting point for a more systematic holographic description of QCD transport.


\subsection{Bulk viscosity}

The bulk viscosity is next analyzed in the non-constant-coupling model. Following the numerical procedure described in Appendix~\ref{bulkviscosity}, the ratio \(\zeta/s\) is computed within the \(O(\alpha)\) approximation for the bulk-channel fluctuation equation. In the present case, the scalar dependence of \(H(\phi)\) modifies the coefficient of the linear \(O(\alpha)\) correction in the fluctuation equation.

\begin{figure}[htbp]
\centering
\includegraphics[width=0.8\textwidth]{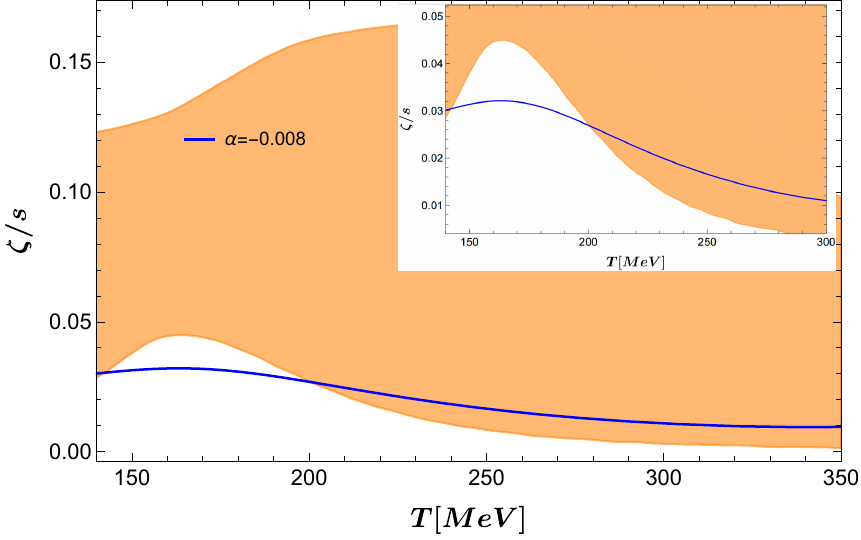}
\caption{\label{fig:zetas_h_nonconstant}
Bulk viscosity to entropy density ratio \(\zeta/s\) as a function of temperature for the non-constant \(H(\phi)\) model with \(\alpha=-0.008\). The blue solid curve denotes the holographic result, while the orange band represents the \(90\%\) credible interval from JETSCAPE~\cite{JETSCAPE:2020mzn}.
}
\end{figure}

Using the parameter set in Table~\ref{tab:parameters2}, the fluctuation equations are solved numerically for
\begin{equation}\label{Hphi}
\alpha=-0.008,\qquad
H(\phi)=1-\frac{\phi^6}{1+2\phi^6}\,,
\end{equation}
which gives the temperature dependence of the bulk viscosity to entropy density ratio shown in Fig.~\ref{fig:zetas_h_nonconstant}. The ratio \(\zeta/s\) develops a pronounced peak near the transition region, with the maximum value
\begin{equation}
\left.\frac{\zeta}{s}\right|_{\rm max}=0.0321
\qquad \text{at} \qquad
T=163.3~\mathrm{MeV}.
\end{equation}
This behavior is qualitatively similar to that found in the constant-coupling case~\eqref{Hconstant} discussed in Section~\ref{bulk viscosity h=1}: the higher-derivative sector changes the magnitude of \(\zeta/s\) while preserving the overall enhancement near the crossover region. The result is broadly compatible with the JETSCAPE phenomenological constraint, shown by the orange band in Fig.~\ref{fig:zetas_h_nonconstant}. Some deviations remain, especially for \(T\in(150,200)~\mathrm{MeV}\), but the model captures the main qualitative behavior of the bulk viscosity once the thermodynamic constraints and the non-monotonic structure of \(\eta/s(T)\) are imposed simultaneously.

Taken together, the results for \(\eta/s\) and \(\zeta/s\) indicate that a scalar-dependent Gauss--Bonnet coupling provides additional flexibility for describing transport properties of strongly coupled QCD matter while maintaining consistency with the equation of state.

\section{Conclusions and Future Perspectives}\label{conclusion}

This work has investigated a higher-derivative extension of holographic QCD obtained by supplementing the EMD framework with a Gauss-Bonnet term coupled to the dilaton. The main objective is to improve the transport sector of holographic QCD while preserving its successful description of equilibrium thermodynamics. After the model parameters are fixed by matching lattice-QCD observables, an adequate description of the equation of state is obtained at both zero and finite baryon chemical potential. For the benchmark choice $H(\phi)=1$, the model breaks the universality of the shear viscosity to entropy density ratio and gives rise to a temperature-dependent $\eta/s$, while also improving the bulk-viscosity sector relative to the two-derivative case. At the same time, a sensible finite-density phase structure is preserved, with the CEP located at $T_C=105~\mathrm{MeV}$ and $\mu_C=556~\mathrm{MeV}$. However, the resulting shear-viscosity profile remains too simple near the crossover region. A qualitatively improved behavior is obtained when the Gauss--Bonnet coupling is allowed to depend nontrivially on the dilaton. For the case in~\eqref{Hphi}, \(\eta/s\) develops a non-monotonic temperature dependence, with a local minimum at \(\left(\left.\eta/s\right|_{\rm min},\, T\right)=(0.0836,\,248.8~\mathrm{MeV})\). By contrast, the temperature dependence of \(\zeta/s\) is already present at the two-derivative level; see, for example, related EMD model studies in~\cite{Grefa:2022sav}. These results indicate that the interplay between higher-curvature corrections and the scalar background can provide a useful mechanism for reconciling holographic transport with lattice-informed thermodynamics within a single framework.

The present analysis is restricted to the specific parameterization given in~\eqref{eq:423}. In the temperature range \(150~\mathrm{MeV}<T<200~\mathrm{MeV}\), the bulk viscosity obtained from the present holographic model shows a visible deviation from the JETSCAPE constraint, as shown in Fig.~\ref{fig:zetas_h_nonconstant}. This deviation may be reduced by adopting a more flexible or better-optimized coupling function. A systematic determination of the corresponding parameter space, however, requires more efficient methods than manual tuning. Machine-learning techniques, such as Gaussian processes, neural networks, or reinforcement learning, provide promising tools for this inverse problem. They could be used to explore the high-dimensional parameter space and optimize the coupling functions directly against multiple lattice and phenomenological observables.

Several directions deserve further investigation. First, the framework can be extended to include additional real-time observables, such as jet-quenching parameters, heavy-quark diffusion coefficients, and electromagnetic correlators, which are closely related to experimental probes in heavy-ion collisions. Second, the near-critical region at finite baryon chemical potential should be examined in more detail, especially the behavior of fluctuations and higher-order susceptibilities associated with criticality. Third, higher-curvature corrections beyond the Gauss-Bonnet term may be incorporated to assess their impact on transport coefficients and their potential role in improving agreement with phenomenological constraints. Finally, more direct constraints from experimental data, such as collective flow and hadron spectra, could be included in the calibration procedure. Such developments would provide a path toward a more systematically constrained and data-driven holographic description of QCD matter.

\acknowledgments

We thank Weijie Fu, Defu Hou, Pak Hang Chris Lau, Weijia Li, and Xiaofeng Luo for useful discussions. 
This work was supported by the National Natural Science Foundation of China under Grants Nos. 12475053, 12525503, 12588101, 12235016, 12447101, and the sub-project funding for “Gravitational Redshift Measurement Scientific Experiment and Frontier Research in Gravitational Physics” of the Chinese Academy of Sciences, the Strategic Priority Research Program on Space Science, the Chinese Academy of Sciences (XDA30040000, XDA30030000).

\appendix
\section{Equations of motion}
\label{app:A}
To derive the Einstein field equations, it is useful to rewrite the Gauss-Bonnet term $R_{GB}^{2}$  as \cite{Myers:1987yn, Deruelle:2018vtt}:
\begin{equation}
    \begin{aligned}
R_{GB}^{2}=R^{\mu\nu\rho\sigma}P_{\mu\nu\rho\sigma}\,,
    \end{aligned}
    \label{eq:413}
\end{equation}
where
\begin{equation}
    \begin{aligned}
{P}_{\;{\,\,\,\,\,\,\rho \sigma }}^{\mu \nu } &= {R}_{\;{\,\,\,\,\,\rho \sigma }}^{\mu \nu } - 2{\delta }_{\lbrack\rho }^{\mu }{R}_{\,\,\,\,\sigma \rbrack }^{\nu } + 2{\delta }_{\lbrack \rho }^{\nu }{R}_{\,\,\,\,\sigma \rbrack }^{\mu } + {\delta }_{\lbrack \rho }^{\mu }{\delta }_{\sigma \rbrack }^{\nu }R \\
 &= \frac{1}{4}{\delta }_{{\rho \sigma }{\beta }_{1}{\beta }_{2}}^{{\mu \nu }{\alpha }_{1}{\alpha }_{2}}{R}_{\;{\,\,\,\,\,\,\,\,\,\,\alpha }_{1}{\alpha }_{2}}^{{\beta }_{1}{\beta }_{2}}\,,
  \end{aligned}
    \label{eq:414}
\end{equation}
here $\delta_{\beta_1\dots\beta_N}^{\alpha_1\dots\alpha_N}$ denotes the generalized Kronecker delta, which is defined as the determinant of the $N\times N$ matrix $M$ constructed from the ordinary Kronecker delta $M_j^i=\delta_{\beta_j}^{\alpha_i}$ \cite{Julie:2019sab}. The quantity \( P_{\mu\nu\rho\sigma} \) has the same symmetry properties as the Riemann tensor and is divergenceless the Bianchi identities immediately give \( \nabla_\mu P^{\mu}_{\,\,\,\ \nu\rho\sigma}=0 \).

The variation of the action \eqref{five-dimensional bulk action} with respect to $g_{\mu\nu}$ can therefore be written as:
\begin{equation}
\begin{aligned}
&R_{\mu\nu}-\frac{1}{2}g_{\mu\nu}R+\alpha\left(H(\phi)\mathcal{H}_{\mu\nu}+4P_{\mu\alpha\mu\beta}\nabla^{\alpha}\nabla^\beta H(\phi)\right)\\
&=-\frac{1}{2}g_{\mu\nu}\Big(\nabla_\mu\nabla^\mu\phi+V(\phi)+\frac{Z(\phi)}{4}F_{\mu\nu}F^{\mu\nu}\Big)+\frac{1}{2}\nabla_\mu\phi\nabla_\nu\phi+\frac{Z(\phi)}{2}F_{\mu\rho}F^{\,\,\,\,\rho}_\nu\,,
    \end{aligned}
    \label{eq:417}
\end{equation}
where $\mathcal{H}_{\mu\nu}$ denotes the Gauss-Bonnet tensor,
\begin{equation}
\begin{aligned}
 \mathcal{H}_{\,\,\,\,\nu }^{\mu }= 2{R}_{\,\,\,\,\alpha \beta \gamma }^{\mu }{P}_{\nu }^{\,\,\,\,\alpha \beta \gamma } - \frac{1}{2}{\delta }_{\nu }^{\mu }{R}^{2}_{\mathrm{{GB}}}=  - \frac{1}{8}{\delta }_{\nu {\beta }_{1}{\beta }_{2}{\beta }_{3}{\beta }_{4}}^{\mu {\alpha }_{1}{\alpha }_{2}{\alpha }_{3}{\alpha }_{4}}{R}_{\;{\,\,\,\,\,\,\,\,\,\,\alpha }_{1}{\alpha }_{2}}^{{\beta }_{1}{\beta }_{2}}{R}_{\;{\,\,\,\,\,\,\,\,\,\,\alpha }_{3}{\alpha }_{4}}^{{\beta }_{3}{\beta }_{4}} .
\end{aligned}
    \label{eq:418}
\end{equation}
In addition, varying the action \eqref{five-dimensional bulk action} with respect to the dilaton field $\phi$, and gauge field $A_\mu$, we can obtain:
\begin{equation}
\begin{aligned}
\nabla_\mu\nabla^\mu\phi-\frac{\partial_\phi Z(\phi)}{4}F_{\mu\nu}F^{\mu\nu}-\partial_\phi V(\phi)+\alpha \partial_{\phi} H(\phi) R_{\mathrm{GB}}^2=0,
    \end{aligned}
    \label{eq:415}
\end{equation}
\begin{equation}
\begin{aligned}
\nabla^\nu\left(Z(\phi)F_{\mu\nu}\right)=0\,.
    \end{aligned}
    \label{eq:416}
\end{equation}

After substituting the metric ansatz~\eqref{eq:metric} into the equations of motion~\eqref{eq:417}, \eqref{eq:415}, and~\eqref{eq:416}, and simplifying the resulting expressions, one obtains the following four independent equations:
\begin{equation}
    \begin{aligned}
	&12 \alpha  r f'' \partial_{\phi}H(\phi )-30 \alpha  r f' \eta ' \partial_{\phi}H(\phi )+24 \alpha  f' \partial_{\phi}H(\phi )-\frac{1}{2} r^3 \eta ' \phi '+3 r^2 \phi '-12 \alpha  r f \eta '' \partial_{\phi}H(\phi )+r^3 \phi ''\\
    &+6 \alpha  r f \eta '^2 \partial_{\phi}H(\phi )-12 \alpha  f \eta ' \partial_{\phi}H(\phi )-\frac{r^3 \partial_{\phi}V(\phi )}{f}\frac{r^3 e^{\eta } A_{t}'^2 \partial_{\phi}Z(\phi )}{2 f}+\frac{12 \alpha  r f'^2 \partial_{\phi}H(\phi )}{f}+\frac{r^3 f' \phi '}{f}=0\,,
    \end{aligned}
    \label{eq:419}
\end{equation}
\begin{equation}
    \begin{aligned}
\frac{r^3 \phi '^2}{2 f}+\frac{3 r^2 \eta '}{2 f}-12 \alpha  r \phi '^2 \partial_{\phi}^{2}H(\phi )-6 \alpha  H(\phi ) \eta ' -\partial_{\phi}H(\phi )\left(18 \alpha  r \eta ' \phi ' +12 \alpha  r \phi '' \right )=0\,,
    \end{aligned}
    \label{eq:420}
\end{equation}
\begin{equation}
    \begin{aligned}
	&\frac{r^3 e^{\eta } A_{t}'^2 Z(\phi )}{4 f^2}
    +\frac{r^3 V(\phi )}{2 f^2}+\frac{3 r}{f}+\frac{r^3 \phi '^2}{4 f}+\frac{3 r^2 f'}{2 f^2}-\frac{18 \alpha  r f' \phi ' \partial_{\phi}H(\phi )}{f}\\
    &-\frac{6 \alpha  f' H(\phi )}{f}-12 \alpha  \phi ' \partial_{\phi}H(\phi )-12 \alpha  r \phi '^2 \partial_{\phi}^{2}H(\phi )-12 \alpha  r \phi '' \partial_{\phi}H(\phi )=0\,,
    \end{aligned}
    \label{eq:422}
\end{equation}
\begin{equation}
    \begin{aligned}
	r \phi '  A_t'  \partial_{\phi}Z(\phi  )+\frac{1}{2} Z(\phi  ) \left(\left(r \eta ' +6\right) A_t' +2 r A_t'' \right)=0\,,
    \end{aligned}
    \label{eq:421}
\end{equation}
where the prime denotes differentiation with respect to \(r\).

To impose regularity at the horizon, the functions \(f(r)\), \(\eta(r)\), \(A_t(r)\), and \(\phi(r)\) are expanded around \(r=r_h\) as
\begin{subequations}
\begin{equation}
f(r) =f_{h}^1\left(r-r_{h}\right)+\cdots \,,
\end{equation}
\begin{equation}
\eta(r)  =\eta_{h}^{0}+\eta_{h}^{1}\left(r-r_{h}\right)+\cdots \,,
\end{equation}
\begin{equation}
A_{t}(r)  =a_{h}^1\left(r-r_{h}\right)+\cdots \,,
\end{equation}
\begin{equation}
\phi(r)  =\phi_{h}^{0}+\phi_{h}^{1}\left(r-r_{h}\right)+\cdots\,,
\end{equation}
    \label{eq:irexpand}
\end{subequations}
substituting the expansion~\eqref{eq:irexpand} into the equations of motion determines the higher-order coefficients recursively. The independent horizon data can be chosen as \(r_h\), \(\eta_h^0\), \(a_h^1\), and \(\phi_h^0\).

An asymptotic expansion near the AdS boundary further shows that the cosmological constant term in \(V(\phi)\) must be corrected. This fixes the parameter \(h_0\) as \(24\alpha\). The resulting near-boundary expansion coefficients are given as follows:
\begin{equation}
\phi(r)=\frac{\phi _s}{r}+\frac{\phi _v}{r^3}-\frac{\ln(r)}{6r^3}\left(\frac{1}{1-4\alpha}-6(c^4_{1}+9c_6^3)\right)+\mathcal{O}\left(\frac{\ln(r)}{r^5}\right)\,,
    \label{eq:429}
\end{equation}
\begin{equation}
\begin{aligned}
A_t(r)=&\mu_B -\frac{2\kappa_N^2n_B}{2r^2}-\frac{2\kappa_N^2n_Bc_3c_5\phi_s}{3(1+c_3)r^3}\\
&+\frac{2\kappa_N^2n_B\phi_s^2\left((1+c_3)^2-6 (1-4 \alpha ) c_3 c_5 ((c_3-1)c_5-6 (c_3+1)c_6)\right)}{48(1+c_3)^2(1-4\alpha)r^4}\\
&-\frac{2\kappa_N^2n_Bc_3c_5^2(1+(-4+c_3)c_3)\phi_s^3}{30(1+c_3)^2r^5}-\frac{2\kappa_N^2 n_Bc_6   \phi_s^3 \left(c_3 ((9(1-4\alpha)c_5^2+2)(1-c_3)+2)+2\right)}{15 (4 \alpha -1) (c_3+1)^2 r^5}\\&-\frac{18\kappa_N^2 n_Bc_3c_5c_6^2\phi_s^3}{25(1+c_3)r^5}+\frac{2\kappa_N^2n_Bc_3c_5((7-12(1-4\alpha)c_1^4)\phi_s^3-60(1-4\alpha)\phi_v)}{300(1+c_3)(1-4\alpha)r^5}\\&-\frac{2\kappa_N^2n_Bc_3c_5\phi_s^3(-1+6(c_1^4+9c_6^2)(1-4\alpha))\ln(r)}{30(1+c_3)(1-4\alpha)r^5}+\mathcal{O}\left(\frac{\ln(r)}{r^6}\right)\,,
\end{aligned}
    \label{eq:430}
\end{equation}
\begin{equation}
\begin{aligned}
\eta(r)=&0+\frac{\phi_s^2}{6r^2(1-4\alpha)}-\frac{4c_6\phi_s^3}{3r^3(1-4\alpha)}\\
&+\frac{\left(1+4\alpha-6c_1^4(1-4\alpha)^2+378 c_6^2 (1 - 4 \alpha)^2\right)\phi^4_s+72(1-4\alpha)^2\phi_s\phi_v}{144r^4(1-4\alpha)^3}\\
&-\frac{\ln(r)}{12r^4(1-4\alpha)^2}\left(1-6(c_1^4+9c_6^2)(1-4\alpha)\right)\phi^4_s\\
&+\frac{36 (1-4 \alpha )^2 c_6 \phi_s^2 \phi_v-2 c_6 \phi_s^5 \left(32 \alpha
   +66 (1-4 \alpha )^2 c_1^4+270 (1-4 \alpha )^2 c_6^2-11\right)}{27 ( 1-4 \alpha)^3 r^5}\\
   &+\frac{2 c_6 \phi_s^5 \ln (r) \left(1-6 (1-4 \alpha ) \left(c_1^4+9 c_6^2\right)\right)}{9
   (1-4 \alpha )^2 r^5}+\mathcal{O}\left(\frac{\ln(r)^2}{r^6}\right),
\end{aligned}
    \label{eq:431}
\end{equation}
\begin{equation}
\begin{aligned}
    f(r)&=r^2+\frac{\phi _s^2}{6(1-4\alpha)}-\frac{4c_6\phi_s^3}{3(1-4\alpha)r}+\frac{f_v}{r^2}-\frac{\ln(r)}{12r^2(1-4\alpha)^2}\left(1-6(c_1^4+9c_6^3)(1-4\alpha)\right)\phi_s^4\\
    &+\frac{4c_6\phi_s^2\left((-7+33c_1^4(1-4\alpha)^2+135c_6^2(1-4\alpha)^2+22\alpha)\phi_s^3-9(1-4\alpha)^2\phi_v\right)}{27(1-4\alpha)^3r^3}\\
    &+\frac{\ln(r)}{9r^3(1-\alpha)^2}\left(1-6(c_1^4+9c_6^3)(1-4\alpha)\right)\phi_s^5+\mathcal{O}\left(\frac{\ln(r)^2}{r^6}\right)\,,
\end{aligned}
    \label{eq:432}
\end{equation}
where \( \mu_B \) and \( n_B \) denote the baryon chemical potential and baryon density, respectively.  \( \phi_s \), which acts as an external source, explicitly breaks conformal symmetry.

\section{Holographic renormalization}\label{app:B}  

To obtain the on-shell action~\eqref{five-dimensional bulk action}, one has to introduce the boundary terms to remove the divergences to render the variational principle well defined~\cite{Skenderis:2002wp,deHaro:2000vlm}. Following~\cite{Davis:2002gn}, the renormalized action supplemented by the required boundary terms takes the form
\begin{equation}
    \begin{aligned}
	S=&\frac{1}{2\kappa_N^2}\int d^5x\sqrt{-g}[R-\frac12\nabla_{\mu}\phi\nabla^{\mu}\phi-V(\phi)-\frac{Z(\phi)}4F_{\mu\nu}F^{\mu\nu}+\alpha H(\phi) R_{GB}^{2} ]\\
    &+\frac{1}{2 \kappa_{N}^2} \int_{\partial_M} d x^{4} \sqrt{-h}(2 K-6+4 \alpha H(\phi) J) \,,      
    \end{aligned}
    \label{eq:actionhk}
\end{equation}
where \(K\) is the trace of the extrinsic curvature, \(K_{ab} = h_{ca}\nabla^{c}n_b\), defined with respect to the normal vector \(n_b\), and \(h_{\mu\nu}\) is the induced metric on the AdS boundary. The quantity \(J\) is the trace of
\begin{equation}
\begin{aligned}
& J_{a b}=\frac{1}{3}\left(2 K K_{a c} K_b^{\,\,\,c}+K_{c d} K^{c d} K_{a b}-2 K_{a c} K^{cd} K_{d b}-K^2 K_{a b}\right).
\end{aligned}
\end{equation}
Substituting the UV expansion into the action~\eqref{eq:actionhk} gives the divergent contribution \(\mathcal{L}_{\text{div}}\):
\begin{equation}
\begin{aligned}
\mathcal{L}_{\text{div}}=&2 (12 \alpha -1) r^4+\frac{(36 \alpha -1) r^2 \phi_s^2}{6-24 \alpha }+\frac{4 (36 \alpha -1) c_6 r \phi_s^3}{12 \alpha -3}\\
&-\frac{(36 \alpha -1) \phi_s^4 \left(6 (4 \alpha -1) c_1^4+54 (4 \alpha -1) c_6^2+1\right) \ln (r)}{12 (1-4 \alpha )^2}\,.
\end{aligned}
    \label{eq:divterm}
\end{equation}
The counterterm Lagrangian is denoted by \(\mathcal{L}_{ct}\). The relevant divergent structures are
    \begin{equation}
\begin{aligned}
  \mathcal{L}_{ct-K}=&\sqrt{-h}K\sim 8r^{4}+\frac{2 r^2 \phi_s^2}{3-12 \alpha }+\frac{16 c_6 r \phi_s^3}{12 \alpha -3}-\frac{\phi_s^4 \left(6 (4 \alpha -1) c_1^4+54 (4 \alpha -1) c_6^2+1\right) \ln (r)}{3 (1-4 \alpha )^2}+\mathcal{O}(1), \\
  \mathcal{L}_{ct-\Lambda}=&\sqrt{-h}\Lambda{\sim}-6r^{4}+\mathcal{O}(1), \\
  \mathcal{L}_{ct-\phi^{2}}=&\sqrt{-h}\phi^{2}\sim r^{2}\phi_{s}^{2}-6 c_6 r \phi_s^3+\frac{\phi_s^4 \left(6 (4 \alpha -1) c_1^4+54 (4 \alpha -1) c_6^2+1\right) \ln (r)}{12 \alpha -3}+\mathcal{O}(1), \\
 \mathcal{L}_{ct-\phi^{3}}= &\sqrt{-h}\phi^{3}\sim r \phi_s^3+\mathcal{O}(r^{-1}), \\
 \mathcal{L}_{ct-\phi^{4}}=& \sqrt{-h}\phi^{4}\sim\phi_{s}^{4}+\mathcal{O}(r^{-1}),\\
  \mathcal{L}_{ct-H(\phi)J}=&\sqrt{-h}H(\phi)J\\
  \sim&-8 r^4+\frac{2 r^2 \phi_s^2}{4 \alpha -1}+\frac{16 c_6 r \phi_s^3}{1-4 \alpha }+\frac{\phi_s^4 \left(6 (4 \alpha -1) c_1^4+54 (4 \alpha -1) c_6^2+1\right) \ln (r)}{(1-4 \alpha )^2}+\mathcal{O}(r^{-1})\,.
\end{aligned}
\label{eq:439}
\end{equation}
Comparison with the divergent term~\eqref{eq:divterm} fixes the boundary action as
\begin{equation}
\begin{aligned}
S_{\partial}=&\frac{1}{2\kappa_{N}^2}\int_{\partial_M}dx^{4}\sqrt{-h}\Big[ (  2K-6+4\alpha H(\phi)J) -\frac{1}{2}\phi^{2}-b\phi^{4}+\frac{1}{4}F_{\rho\lambda}F^{\rho\lambda}\ln(r)-c_6 \phi^{3} \\
 &-\left(\frac{\alpha }{12 \alpha -3}+\frac{9 c_6^2}{2}+\frac{1}{12} \left(6 c_1^4-1\right)\right)\phi^{4}\ln(r)\Big ]\,.
\end{aligned}
    \label{eq:action_boundary}
\end{equation}

Following the \cite{Davis:2002gn} and combined with the results of the counterterms, the Brown-York stress tensor on the boundary is given by:
\begin{equation}
\begin{aligned}
T_{ab} =&\operatorname*{lim}_{r\to\infty}\frac{2r^{2}}{\sqrt{-\det h}}\frac{\delta\left(S+S_{\partial}\right)_{\mathrm{on-shell}}}{\delta h^{ab}} \\
  =&\frac{1}{2\kappa_{N}^2}\operatorname*{lim}_{r\to\infty}r^{2}\Big[2\left(Kh_{ab}-K_{ab}-3h_{ab}\right)  +4\alpha H(\phi)(3J_{ab}-Jh_{ab}) \\
&+4\alpha\Big(2K_{ea}K_{b}^{\,\,\,e}-2KK_{ab}+h_{ab}(K^{2}-K_{cd}K^{cd})\Big)n^{e}\partial_{e}H(\phi) \\
&-\left(\frac{1}{2}\phi^2+b\phi^4-c_6 \phi^3\right.   \left.+\left(\frac{\alpha }{12 \alpha -3}+\frac{9 c_6^2}{2}+\frac{1}{12} \left(6 c_1^4-1\right)\right)\phi^4\ln(r) \right)h_{ab}\\
&-\left(F_{ac}F_{b}^{\,\,\,c}-\frac{1}{4}h_{ab}F_{cd}F^{cd}\right)\ln(r)\Big].
\end{aligned}
    \label{eq:Energy-Momentum Tensor}
\end{equation}
The components of the Brown-York stress tensor correspond to the thermodynamic quantities in the field theory. By substituting the UV expansion into~\eqref{eq:Energy-Momentum Tensor} and expanding at the boundary, we obtain the energy density and pressure:
\begin{equation}
\begin{aligned}
\epsilon&=T_{tt}\\&
=\frac{1}{2\kappa_{N}^2}\Big(\frac{\alpha  \phi _s^4}{12 (1-4 \alpha )^2}+\frac{\phi _s^4}{48 (1-4 \alpha )^2}+b \phi _s^4+\frac{27}{2} c_6^2 \phi _s^4+3 (4 \alpha -1) f_s+\phi _s \phi _v\Big),
\end{aligned}
    \label{eq:442}
\end{equation}
\begin{equation}
\begin{aligned}
P&=T_{xx}=T_{yy}=T_{zz}\\&=
\frac{1}{2\kappa_{N}^2}\Big((4 \alpha -1) f_v+\phi s \phi _v+3 c_6^2 \phi _s^4-b \phi _s^4-\frac{\phi _s^4 \left(28 \alpha +24 (1-4 \alpha )^2 c_1^4-9\right)}{144 (1-4 \alpha )^2}\Big),
\end{aligned}
    \label{eq:443}
\end{equation}
The trace anomaly is obtained by computing the trace of the stress tensor, which is given by 
\begin{equation}
\begin{aligned}
\epsilon-3P=&\frac{1}{2\kappa_{N}^2}\Big(\frac{2 \alpha  \phi_s^4}{3 (1-4 \alpha )^2}-\frac{\phi_s^4}{6 (1-4 \alpha )^2}+4 b \phi_s^4+\frac{1}{2} c_1^4 \phi_s^4+\frac{45}{2} c_6^2 \phi_s^4-2 \phi_s \phi_v\Big).
\end{aligned}
    \label{eq:444}
\end{equation}
The free energy density can be calculated via the following expression:
\begin{equation}
\begin{aligned}
-\Omega V=T\left(S+S_{\partial}\right)_{\text{on-shell}},
\end{aligned}
    \label{eq:445}
\end{equation}
where \( V = \int dx\,dy\,dz \) and \( t \in [0, 1/T] \). From the equations of motion \eqref{eq:419}-\eqref{eq:421} and the asymptotic expansion~\eqref{eq:429}-\eqref{eq:432}, we obtain
\begin{equation}
\begin{aligned}
\Omega=&\frac{1}{2\kappa_{N}^2}\operatorname*{lim}_{r\to\infty}\Big(2e^{-\eta/2}r^{2}f+8\alpha e^{-\eta/2}f(-rf^{\prime}+f(r\eta^{\prime}-1))(H(\phi)-r\partial_\phi H(\phi)\phi^{\prime})\Big) \\
&-e^{-\eta/2}r^{3}\sqrt{f}\Bigg((2K-6+4\alpha H(\phi)J)-\frac{1}{2}\phi^{2} -c_6\phi^{3}-b\phi_s^4-\left(\frac{\alpha }{12 \alpha -3}+\frac{9 c_6^2}{2}+\frac{1}{12} \left(6 c_1^4-1\right)\right)\phi^{4}\ln(r)\Bigg)\\
=&-\frac{1}{2\kappa_{N}^2}\Big((4 \alpha -1) f_v+\phi_s \phi_v+3 c_6^2 \phi_s^4-b \phi_s^4-\frac{\phi_s^4 \left(28 \alpha +24 (1-4 \alpha )^2 c_1^4-9\right)}{144 (1-4 \alpha )^2}\Big).
\end{aligned}
    \label{eq:446}
\end{equation}
A comparative analysis reveals that the relationship between the free energy and pressure remains fully consistent with thermodynamic relations.
\begin{equation}
    \begin{aligned}
 P = {sT} + {\mu_B n_B } - \epsilon 
 =  - \Omega .
 \end{aligned}
    \label{eq:47}
\end{equation}
Here, the black hole entropy is computed via the Wald formula, 
\begin{equation}  
\begin{aligned} 
s=-2\pi\oint_{\Sigma}d^3x\sqrt{-h}\frac{\delta\mathcal{L}}{\delta R_{\mu\nu\rho\sigma}}\epsilon_{\mu\nu}\epsilon_{\rho\sigma}, 
\end{aligned}  
    \label{eq:472}  
\end{equation}  
which yields 
\begin{equation}  
\begin{aligned}  
s=\frac{2\pi}{\kappa_{N}^2 }r_{h}^{3}\,.  
\end{aligned}  
\label{eq:Tands}  
\end{equation} 
In addition, a useful conserved charge $Q$ of the system through the equations of motion is given by:  
\begin{equation}  
\begin{aligned}  
Q=\frac{1}{2\kappa_N^2}r^3e^{\eta/2}\left(\left(r^2+4\alpha fH(\phi)+8\alpha rf\partial_\phi H(\phi)\phi'\right)\left(\frac{f}{r^2}e^{-\eta}\right)'-Z(\phi)A_tA'_t\right).  
\end{aligned}  
    \label{eq:449}  
\end{equation}  
Comparing with the model \cite{Cai:2022omk}, our conserved charge incorporates modifications from the Gauss-Bonnet term. The modified charge still maintains consistency between horizon and boundary expressions. Substituting the IR expansion, we obtain:  
\begin{equation}  
\begin{aligned}  
Q=Ts\,. 
\end{aligned}  
    \label{eq:450}  
\end{equation}  
At the AdS boundary, substituting the UV expansion yields:  
\begin{equation}  
\begin{aligned}  
Q=\epsilon+P-\mu_B n_B\,.
\end{aligned}  
    \label{eq:451}  
\end{equation}

\section{Calculation of shear and bulk viscosities}
\label{app:viscosity} 

This section presents the technical details of the holographic computation of the shear and bulk viscosities.

\subsection{Shear viscosity}
\label{app:C}
we start from a general ansatz for the background metric:
\begin{equation}
\label{ansatzmetric}
    \mathrm{d}s^2 = -C_1(r)^2 \mathrm{d}t^2  + C_3(r)^2 \mathrm{d}r^2+ C_2(r)^2 (dx^2+dy^2+dz^2).
\end{equation}
The transport coefficients of finite temperature plasma can be extracted through various approaches. For the shear viscosity, the most straightforward approach stems from the Kubo formula \cite{Myers:2009ij}:
\begin{equation}
\begin{aligned}
 \eta  =  - \mathop{\lim }\limits_{{\omega  \rightarrow  0}}\frac{1}{\omega }\operatorname{Im}{G}_{{xy},{xy}}^{R}\left( {\omega ,,\mathbf{k} = 0}\right) , 
\end{aligned}
    \label{eq:456}
\end{equation}
where the retarded Green's function is  defined as:
\begin{equation}
\begin{aligned}
 {G}_{{xy},{xy}}^{R}\left( {\omega ,\mathbf{k} = 0}\right)  =  - i\int {dtd}\mathbf{x}{e}^{i\omega t}\theta \left( t\right) \left\langle  \left\lbrack  {{T}_{xy}\left( x\right) ,{T}_{xy}\left( 0\right) }\right\rbrack  \right\rangle  . 
\end{aligned}
    \label{eq:457}
\end{equation}
Within the AdS/CFT correspondence, the retarded Green's function is computed by introducing small metric perturbations in the dual gravitational theory, we should add a small bulk perturbation
${h}_{xy}\left( {t,r}\right)$ to the metric~\eqref{ansatzmetric}. The calculation starts from the effective action for {the perturbation \(h_x^{\,\,y}(t,r)\),} whose Fourier decomposition is
\begin{equation}
\begin{aligned}
h_x^{\,\,y}(t,r) = \int \frac{{d}^{4}k}{{\left( 2\pi \right) }^{4}}{\phi }_{k}\left( r\right) {e}^{-{i\omega t} + {ikz}}. 
\end{aligned}
    \label{eq:458}
\end{equation}
Expanding the action~\eqref{five-dimensional bulk action} to quadratic order in the fluctuation \(\phi_k(u)\) gives
\begin{equation}
\begin{aligned}
& I_\phi^{(2)}=\frac{1}{2 \kappa_N^2} \int \frac{d^4 k}{(2 \pi)^4} d r\left(A(r) \phi_k^{\prime \prime} \phi_{-k}+B(r) \phi_k^{\prime} \phi_{-k}^{\prime}+C(r) \phi_k^{\prime} \phi_{-k}\right. \\
&\left.+D(r) \phi_k \phi_{-k}+E(r) \phi_k^{\prime \prime} \phi_{-k}^{\prime \prime}+F(r) \phi_k^{\prime \prime} \phi_{-k}^{\prime}\right)+\mathcal{K} \,,
\end{aligned}
\end{equation}
where \(\mathcal{K}\) denotes the generalized Gibbons-Hawking boundary term, which is introduced to solve the variational problem~\cite{Buchel:2004di}. As discussed in \cite{Myers:2009ij}, when calculating the shear viscosity $\eta$ in the low-frequency limit, the only relevant contribution in the flux arises from the canonical momentum term. The shear viscosity $\eta$  reduces to:
\begin{equation}
\begin{aligned}
\eta  = & - \mathop{\lim }\limits_{{\omega  \rightarrow  0}}\frac{1}{\omega }\operatorname{Im}{G}_{{xy},{xy}}^{R}\left( {\omega ,\mathbf{k} = 0}\right)  = \frac{1}{\kappa_N^2}\left( {{\kappa }_{2}\left( {r}_{h}\right)  + {\kappa }_{4}\left( {r}_{h}\right) }\right) \, .
\end{aligned}
    \label{eq:469}
\end{equation}
The functions $\kappa_2(r)$ and $\kappa_4(r)$ are defined as:
\begin{equation}
\begin{aligned}
{\kappa }_{2}\left( r\right) & = \sqrt{-\frac{C_3\left( r\right) }{C_1\left( r\right) }}\left( {A\left( r\right)  - B\left( r\right)  + \frac{{F}^{\prime }\left( r\right) }{2}}\right) ,\ \ \ \ 
{\kappa }_{4}\left( r\right)  = {\left( E\left( r\right) {\left( \sqrt{-\frac{C_3\left( r\right) }{C_1\left( r\right) }}\right) }^{\prime }\right) }^{\prime }\,.
\end{aligned}
    \label{eq:470}
\end{equation}
For the metric~\eqref{eq:metric}, the functions $A(r)$, $B(r)$, $E(r)$, and $F(r)$ are given by:
    \begin{equation}
\begin{aligned}
\frac{A(r)}{\sqrt{-g}}=&2 r f(r) \sqrt{e^{-\eta (r)}} \left(r^2-4 \alpha  H(\phi (r)) \left(r f'(r)-r f(r) \eta '(r)+f(r)\right)\right),\\
\frac{B(r)}{\sqrt{-g}}=&-\frac{2 \alpha  r^2 f'(r)^2 H(\phi (r))+r f(r) \left(\alpha  H(\phi (r)) \left(2 r f''(r)+f'(r) \left(28-5 r \eta '(r)\right)\right)-3 r\right)}{2 r^2}\\
&-\frac{\alpha  f(r)^2 H(\phi (r)) \left(-2 r^2 \eta ''(r)+r^2 \eta '(r)^2-22 r \eta '(r)+24\right)}{2 r^2},\\
F(r)=&\frac{2 \alpha  f(r) H(\phi (r)) \left(f(r) \left(r \eta '(r)-2\right)-r f'(r)\right)}{r},\\
E(r)=&0\,.
\end{aligned}
    \label{eq:471}
\end{equation}
With these expression \eqref{eq:469}, \eqref{eq:470} and \eqref{eq:471}, it is straightforward to obtain the shear viscosity:

    \begin{equation}
\begin{aligned}
\eta=\frac{r_h^2}{2 \kappa_{N}^2}\big(r_h-2 \alpha  r_h f'\left(r_h\right) \phi '\left(r_h\right) \partial_{\phi} H\left(\phi \left(r_h\right)\right)-2 \alpha  f'\left(r_h\right) H\left(\phi \left(r_h\right)\right)\big)\,.
\end{aligned}
    \label{eq:472}
\end{equation}
Using~\eqref{eq:Tands}, the ratio $\eta/s$ ultimately reduces to:
\begin{equation}
\begin{aligned}
 \frac{{\eta }}{s} = \frac{1}{4\pi }\Big[1- \frac{2 \alpha}{r_h} f'(r_h) \big(\phi '(r_h) \partial_{\phi} H(\phi (r_h))+H(\phi (r_h))\big)\Big].
\end{aligned}
    \label{eq:473}
\end{equation}

\subsection{Bulk viscosity} 
\label{bulkviscosity}

This section provides a detailed derivation of the bulk-viscosity formula~\eqref{eq:zetasfinal_main}. Following~\cite{Buchel:2023fst,Tong:2025rxz}, the ansatz~\eqref{ansatzmetric} is substituted into the equations of motion~\eqref{eq:415}, \eqref{eq:416}, and~\eqref{eq:417}, yielding
\begin{align}
\frac{C_1'}{C_2} - \frac{C_2' C_3'}{C_3} + \frac{C_2'^2}{C_2} + \frac{1}{12} C_2 \phi'^2 + \frac{1}{6} C_2 C_3^2 V(\phi) + \frac{C_2 Z(\phi)}{12 C_1^2} A_{t}'^2 + \alpha \cdot [\cdots] &= 0,  \\
\phi'^2 - \frac{12 C_2'^2}{C_2^2} - \frac{12 C_1' c_2'}{C_1 C_2} - 2 C_3^2 V(\phi) - \frac{Z(\phi)}{C_1^2} A_{t}'^2 + \alpha \cdot [\cdots] &= 0,  \\
C_1'' + 3 \frac{C_1' C_2'}{c_2} - \frac{C_1' C_3'}{C_3} + \frac{1}{3} C_1 C_3^2 V(\phi) - \frac{Z(\phi) A_{t}'^2}{3 C_1} + \alpha \cdot [\cdots] &= 0,  \\
A_{t}'' - \frac{C_1'}{C_1} A_{t}' + 3 \frac{C_2'}{C_2} A_{t}' - \frac{C_3'}{C_3} A_{t}' + \frac{\partial_{\phi} Z(\phi)}{Z(\phi)} A_{t}' &= 0,  \\
\phi'' + \frac{C_1'}{C_1} \phi' + 3 \frac{C_2'}{C_2} \phi' - \frac{C_3'}{C_3} \phi' - C_3^2 \partial_{\phi} V(\phi) + \frac{\partial_{\phi} Z(\phi)}{2 C_1^2} A_{t}'^2 + \alpha \cdot [\cdots] &= 0, 
\end{align}
where the omitted terms $\alpha \cdot [\cdots]$ represent the contribution of higher-order curvature derivatives, due to its complexity, we do not provide the details.

To compute the bulk viscosity, consider the decoupled set of SO(3)-invariant metric fluctuations together with the bulk scalar fluctuation,
\begin{equation}
    \mathrm{d}s_{5}^2 \to \mathrm{d}s_{5}^2 +h_{tt}(t,r) \mathrm{d}t^2 + h_{11}(t,r)(dx^2+dy^2+dz^2), \quad \phi \to \phi + \psi (t,r).
\end{equation}
For convenience, the axial gauge is imposed,
\begin{equation}
h_{tr} = h_{rr} = 0 .
\end{equation}
 The metric perturbations can be rewritten as
 \begin{equation}
     h_{tt}(t,r) = e^{-i \omega t} C_{1}^{2}H_{00}(r), \quad h_{11}(t,r) = e^{-i \omega t} C_{2}^{2}H_{11}(r),
 \end{equation}
 and the gauge invariant scalar fluctuations are given by \cite{Benincasa:2005iv}
 \begin{equation}
     \psi(t,r) = e^{-i \omega t}\left(Z_{\phi}(r)+\frac{\phi'C_{2}}{2C_{2}'}H_{11}(r)\right).
     \label{psidef}
 \end{equation} 
Substituting these fluctuations into the equations of motion and linearizing around the background yields the following coupled linear equations:
 \begin{equation}
     \begin{aligned}
0= & Z_{\phi}^{\prime \prime}+\left(\ln \frac{C_{1} C_{2}^{3}}{C_{3}}\right)^{\prime} Z_{\phi}^{\prime}-Z_{\phi}\left(\frac{4 C_{2} C_{3}^{2} C_{1}^{\prime} V(\phi)}{3 C_{1} C_{2}^{\prime}}+\frac{4}{3} C_{3}^{2} V(\phi)+\frac{2 C_{2}^{2} C_{3}^{4} V(\phi)^{2}}{9 C_{2}^{\prime 2}}+
C_{3}^{2} \partial^2_{\phi}V(\phi) \right.\\
&\left.+\frac{2 C_{2} C_{3}^{2} \phi^{\prime} \partial_{\phi}V(\phi)}{3 C_{2}^{\prime}}\right)+Z_{\phi}\left(-\frac{2 C_{2}^{2} C_{3}^{2} V(\phi) Z(\phi)\left(A_{t}^{\prime}\right)^{2}}{9 C_{1}^{2}\left(C_{2}^{\prime}\right)^{2}}+\frac{\left(A_{t}^{\prime}\right)^{2} \partial^2_{\phi}Z(\phi)}{2 C_{1}^{2}}-\frac{C_{2}^{2} Z(\phi)^{2}\left(A_{t}^{\prime}\right)^{4}}{18 C_{1}^{4}\left(C_{2}^{\prime}\right)^{2}}\right. \\
& \left.+\frac{C_{2}\left(A_{t}^{\prime}\right)^{2} \partial_{\phi}Z(\phi)\phi^{\prime}}{3 C_{1}^{2} C_{2}^{\prime}}-\frac{\left(A_{t}^{\prime}\right)^{2}\left(\partial_{\phi}Z(\phi)\right)^{2}}{C_{1}^{2} Z(\phi)}-\frac{2 C_{2} Z(\phi)\left(A_{t}^{\prime}\right)^{2} C_{1}^{\prime}}{3 C_{1}^{3} C_{2}^{\prime}}-\frac{2  Z(\phi)\left(A_{t}^{\prime}\right)^{2}}{3 C_{1}^{2}}\right)+\frac{C_{3}^{2} \omega^{2}}{C_{1}^{2}} Z_{\phi}+\alpha \cdot[\cdots]\,,
\end{aligned}
\label{Zphiequation}
 \end{equation}
 and 
\begin{equation}
\begin{aligned}
0=&H_{00}+\frac{2 Z_{\phi} Z\partial_{\phi}Z(\phi)}{Z(\phi)}+\frac{\bar{H}\left(3 C_{2}^{\prime}Z(\phi)+C_{2} \phi^{\prime} \partial_{\phi}Z(\phi)\right)}{C_{2} C_{3} Z(\phi)}\,, \\
0=&\bar{H}^{\prime}+\frac{C_{3} C_{2}}{3C_{2}^{\prime}}Z_{\phi}\phi^{\prime}+\alpha \cdot[\cdots] \,,
\end{aligned}
 \end{equation}
 where 
 \begin{equation}
 \bar{H}\equiv\frac{C_{2} C_{3}}{C_{2}^{\prime}} H_{11}.
 \end{equation}
In addition, two points should be noted:
\begin{itemize}
    \item In~\eqref{Zphiequation}, the \(\alpha\)-dependent contribution is associated only with the function \(Z_{\phi}\). For \(H(\phi)=1\), the equation for \(Z_{\phi}\) can be determined explicitly, in analogy with the results of~\cite{Buchel:2024umq}. When \(H(\phi)\) depends on \(\phi\), however, the resulting equation becomes considerably more involved. In the present derivation, only terms up to \(O(\alpha)\) are retained.
    \item In the near-boundary expansion of \(\psi\), the non-normalizable mode must vanish~\cite{Benincasa:2005iv}. As follows from~\eqref{psidef}, \(Z_{\phi}\) is closely related to \(\phi\) near the AdS boundary. This relation fixes the structure of the UV expansion of \(Z_{\phi}\), which will be used in the subsequent numerical analysis.
\end{itemize}
We rewrite the perturbations as
\begin{equation}
h_{t t}(t, r)=e^{-i \omega t} h_{00, \omega}(r), \quad h_{11}(t, r)=e^{-i \omega t} h_{11, \omega}(r), \quad \psi(t, r)=e^{-i \omega t} p_\omega(r) .
\end{equation}
The effective action at quadratic order in  perturbations can be expressed as
\begin{equation}
\begin{aligned}
S^{(2)}=\frac{1}{2\kappa_N^2} \int d r \left(h_{00,  \omega}^* \cdot \frac{\delta S^{(2)}}{\delta h_{00, \omega}^*}+h_{11, \omega}^* \cdot \frac{\delta S^{(2)}}{\delta h_{11, \omega}^*}+p_\omega^* \cdot \frac{\delta S^{(2)}}{\delta p_\omega^*}+\partial_r J_\omega \right),
\end{aligned}
\end{equation}
where
\begin{equation}
\begin{aligned}
J_\omega
= \frac{1}{12 C_1^4 C_2^2 C_3}\Biggl[
&9 C_2^5 h_{00,\omega}^* h_{00,\omega} C_1^{\prime}
-9 C_1^2 C_2^3 h_{00,\omega}^* h_{11,\omega} C_1^{\prime}
-3 C_1 C_2^5\left(
h_{00,\omega} h_{00,\omega}^{* \prime}
+h_{00,\omega}^* h_{00,\omega}^{\prime}
\right)
\\
&+C_1^5\left(
9 C_2 h_{11,\omega} h_{11,\omega}^{* \prime}
+9 h_{11,\omega}^*
\left(h_{11,\omega} C_2^{\prime}
-C_2 h_{11,\omega}^{\prime}\right)
+16 C_2^5 p_\omega^* p_\omega^{\prime}
+24 C_2^3 p_\omega^* h_{11,\omega} \phi^{\prime}
\right)
\\
&+C_1^3 C_2^2\left(
-9 h_{11,\omega}^* h_{00,\omega} C_2^{\prime}
+C_2\left(
9 h_{00,\omega} h_{11,\omega}^{* \prime}
+9 h_{11,\omega} h_{00,\omega}^{* \prime}
+8 C_2^2 p_\omega^* h_{00,\omega} \phi^{\prime}
\right)
\right)
\Biggr]\\
&+\alpha \cdot[\cdots]\, .
\end{aligned}
\label{boundaryJw}
\end{equation}
The (\ref{boundaryJw}) implies that the imaginary part of $J_\omega$ remains constant along a radial direction.

As $\mathfrak{w}=\omega/2 \pi T \ll 1$, the fluctuations $Z_{\phi}$, $\bar{H}$ and $H_{00}$ can be expressed as
\begin{equation}
Z_\phi=\left(\frac{C_1}{C_2}\right)^{-\imath \mathfrak{w}}\left(z_0+i \mathfrak{w} z_1\right), \quad  \bar{H}=H_0+i \mathfrak{w} H_1\,, \quad H_{00}=H_{00,0}+i \mathfrak{w} H_{00,1}\,,
\label{tranformZHH00}
\end{equation}
and also satisfy the boundary conditions $\lim _{r \rightarrow \infty} H_{11}(r)=1, \lim _{r \rightarrow \infty} H_{00}(r)=0$.

From the definition of the Kubo relation 
\begin{equation}
\zeta=-\frac{4}{9} \lim _{\omega \rightarrow 0} \frac{1}{\omega} \operatorname{Im} G_R=-\frac{4}{9\kappa^2_N} \lim _{\omega \rightarrow 0} \frac{1}{\omega} \operatorname{Im} J_\omega,
\end{equation}
we can get the bulk viscosity
\begin{equation}
\label{bulkzetas1}
    \zeta=\frac{s}{4\pi} \cdot \frac{4}{9} (z_0^{h})^2,
\end{equation}
where $z_0^{h}$ is the value of $Z_{\phi}$ at the zero frequency on the horizon. Therefore, the bulk viscosity to entropy density ratio  is given by
\begin{equation}
\label{bulkzetas2}
    \frac{\zeta}{s}=\frac{1}{4\pi} \cdot \frac{4}{9}(z_0^{h})^2\,.
\end{equation}
We arrive at the result shown in (\ref{eq:zetasfinal_main}).
 Deriving the final results from the formula (\ref{bulkzetas2}) necessitates substantial numerical work. 
First, we set the general ansatz (\ref{ansatzmetric}) as follows
\begin{equation}
     C_{1} \rightarrow \sqrt{f(r)} e^{-\frac{\eta(r)}{2} }, \quad C_{2} \rightarrow r, \quad C_{3} \rightarrow \frac{1}{\sqrt{f(r)}}\,,
\end{equation}
which transforms the metric into our ansatz (\ref{eq:metric}). From the (\ref{Zphiequation}) and (\ref{tranformZHH00}), we can get the equation of motion $z_{0}$:
\begin{equation}
\begin{aligned}
0 = &z_{0}'' + \left(\frac{f'}{f}-\frac{r\eta'-6}{2r}\right)z_{0}' +\biggl\{ -r(2V(\phi)+e^{\eta}Z(\phi) A_{t}'^{2})(2rV(\phi)+ re^{\eta}Z(\phi) A_{t}'^{2}+6f') \\
&+ 3f\biggl(4V(\phi)(r\eta'-2)-4r\partial_{\phi}V(\phi)\phi' + \frac{e^{\eta}A_{t}'^{2}}{Z(\phi)}\biggl(-6(\partial_{\phi}Z(\phi))^{2} \\&+ Z(\phi)\Bigl(2Z(\phi)\left(r\eta'-2\right) + 2r\partial_{\phi}Z(\phi)\phi' + 3\partial^2_{\phi}Z(\phi)\Bigr)\biggr)- 6\partial^2_{\phi}V(\phi) \biggr) \biggr\} \frac{z_{0}}{18f^2}   + \alpha\cdot[\cdots].
\end{aligned}
\label{zeom}
\end{equation}

To obtain the bulk viscosity, we numerically solve the system comprised of (\ref{zeom}). It is crucial to note that the value of $\psi$ vanishes at the AdS boundary. From (\ref{psidef}), this implies that, under the UV expansion, we have
\begin{equation}
    \lim_{r \to \infty }z_{0 }\cdot r =\frac{1}{2}\phi _{s} ,
    \label{zuvcondition}
\end{equation}
which establishes the relation between $\phi$ and  of $z_{0}$ at the AdS boundary. Therefore, the UV expansion for $z_{0}$ can be readily formulated\footnote{The result (\ref{zuvexpansion}) holds regardless of whether $H(\phi)$ is a constant or not. This is because $\lim_{r \to \infty}H(\phi)=1$ at the UV boundary.}
\begin{equation}
  z_{0}(r)=  \frac{\phi_ s}{2 r}+\frac{z_ v}{r^3}+\frac{\ln(r)}{4r^3}(6c_1^4-1-4\alpha)\phi_s^3+\mathcal{O}\left(\frac{\ln (r)}{r^5}\right),
  \label{zuvexpansion}
\end{equation}
where $c_1$ is the parameter mentioned in function $V(\phi)$, and we have already set $H(\phi)$ in the form as shown in (\ref{eq:423}).

The horizon regularity of $z_{0}$ ensures that it is described by an expansion of the form
\begin{equation}
    z_{0}=z_{h}^0+z_{h}^1(r-r_{h})+\cdots\,\,.
\end{equation}
Using the UV and IR boundary conditions together with the equations of motion, \(z_0\) can be solved numerically, and its horizon value is obtained as \(z_0^h=z_0(r_h)\).

 \bibliographystyle{JHEP}
 \bibliography{Newbib.bib}
\end{document}